\documentclass[10pt]{iopart}

\usepackage{iopams}

\expandafter\let\csname equation*\endcsname\relax
\expandafter\let\csname endequation*\endcsname\relax
\usepackage{physics}

\usepackage{amsfonts}
\usepackage{mathtools}
\usepackage{algorithm}
\usepackage{algpseudocode}
\usepackage{latexsym}

\usepackage{url}
\usepackage[dvipsnames]{xcolor}

\usepackage{mleftright}

\usepackage{graphicx}
\usepackage{caption}
\usepackage{subcaption}

\usepackage{tikz}
\usetikzlibrary{plotmarks}
\usepackage{pgfplots}
\pgfplotsset{compat=1.12}
\usepackage{float}
\usepackage[]{placeins} %
\usepackage{enumitem}
\usepackage{xifthen}
\usepgfplotslibrary{external}
\usepgfplotslibrary{fillbetween}
[
prefix={figures/}
]
\pgfplotsset{
	every tick label/.append style={font=\small},
	every axis/.append style={
		width = 0.45\textwidth,
		mark size = 2.85pt,
		axis line style = ultra thick,
		legend cell align={left}, 
		legend style={draw=none},
		legend style={font=\small},
		legend style={fill=none, text opacity=1}
	},
    every axis plot/.append style={very thick} 
}
\usepackage{pdfpages}
\usepackage{booktabs}
\usepackage[toc,page]{appendix}
\usepackage{float}
\usepackage{slashbox}
\usepackage{subcaption}

\DeclarePairedDelimiterX{\mean}[1]{\langle}{\rangle}{#1}

\newcommand{\Lorentz}{\mathcal{L}}
\newcommand{\MONKES}{{\texttt{MONKES}}}
\newcommand{\DKES}{{\texttt{DKES}}}
\newcommand{\ii}{\text{i}}
\newcommand{\Matrix}[2]
{
	\mleft[
	\begin{array}{#1}
		#2
	\end{array}
	\mright]
}

\begin{document}

\title[MONKES: a fast neoclassical code for the evaluation of monoenergetic transport coefficients]{MONKES: a fast neoclassical code for the evaluation of monoenergetic transport coefficients}

\author{F. J. Escoto$^1$, J. L. Velasco$^1$, I. Calvo$^1$, M. Landreman$^2$ and F. I. Parra$^3$}
\address{$^1$Laboratorio Nacional de Fusión, CIEMAT, 28040 Madrid, Spain}
\address{$^2$University of Maryland, College Park, MD 20742, USA}
\address{$^3$Princeton Plasma Physics Laboratory, Princeton, NJ 08540, USA}
\ead{fjavier.escoto@ciemat.es}
\vspace{10pt}
\begin{indented}
\item[]April 2023
\end{indented}

\begin{abstract}
	{\MONKES} is a new neoclassical code for the evaluation of monoenergetic transport coefficients in stellarators. By means of a convergence study and benchmarks with other codes, it is shown that {\MONKES} is accurate and efficient. The combination of spectral discretization in spatial and velocity coordinates with block sparsity allows {\MONKES} to compute monoenergetic coefficients at low collisionality, in a single core, in approximately one minute. {\MONKES} is sufficiently fast to be integrated into stellarator optimization codes for direct optimization of the bootstrap current and to be included in predictive transport suites. The code and data from this paper are available at \texttt{https://github.com/JavierEscoto/MONKES/}.   
\end{abstract}

%
\vspace{2pc}
\submitto{\NF}
%
%
\ioptwocol

\section{Introduction}
\label{sec:Introduction}
Stellarators are an attractive alternative to tokamaks as future fusion reactors. While tokamaks require a large toroidal current to generate part of the magnetic field, in stellarators the field is produced entirely by external magnets. As a consequence, stellarators avoid current-induced instabilities and facilitate steady-state operation. These advantages come at the expense of making the magnetic field three-dimensional. In tokamaks, axisymmetry guarantees that the radial displacement that charged particles experience along their collisionless orbits averages to zero. Therefore, in the absence of collisions, all charged particles are confined. However, in a generic stellarator the orbit-averaged radial drift velocity does not vanish for trapped particles and they quickly drift out of the device. The combination of a non zero orbit-averaged radial drift and a small collision frequency (reactor-relevant fusion plasmas are weakly collisional in the core) produces, for a generic stellarator, intolerably large levels of neoclassical transport. 

Hence, stellarator magnetic fields must be carefully designed in order to display good confinement properties. This process of tailoring of the magnetic field is called stellarator optimization. The objective of neoclassical optimization is to obtain a stellarator with levels of neoclassical losses equivalent or lower to those in an axisymmetric device. Stellarator magnetic fields in which the orbit-averaged radial magnetic drift is zero for all particles are called omnigenous \cite{Cary1997OmnigenityAQ}. Thus, the goal of neoclassical optimization is to obtain magnetic fields which are close to omnigeneity. However, addressing only radial transport in the optimization process is not sufficient. In toroidal plasmas, the parallel flow of electrons and the rest of species is not, in general, balanced. This mismatch produces a net parallel current at each flux surface which, through Ampère's law, modifies the magnetic field $\vb*{B}$. When the current is generated by a combination of neoclassical mechanisms and non-zero plasma profile gradients, we speak of bootstrap current. The bootstrap current and its effect on the magnetic configuration must be taken into account in the design of optimized stellarator magnetic fields. 

Two different subclasses of omnigenous stellarators have drawn particular attention: quasi-isodynamic (QI) and quasi-symmetric (QS) stellarators. Quasi-isodynamic configurations are omnigeneous configurations in which the curves of constant magnetic field strength $B:=|\vb*{B}|$ on a flux surface close poloidally. This additional property has an important implication: exactly QI stellarators produce zero bootstrap current at low collisionality \cite{Helander_2009,Helander_2011_Bootstrap}. Thanks to this feature, QI stellarators can control plasma-wall interaction by means of a divertor relying on a specific structure of islands, which could not be realized in the presence of large toroidal currents. The Wendelstein 7-X (W7-X) experiment was designed to be close to QI and demonstrates that theoretically based stellarator optimization can be applied to construct a device with much better confinement properties than any previous stellarator \cite{Beidler2021}. Moreover, the bootstrap current produced in W7-X plasmas is smaller than in non-optimized machines \cite{Dinklage2018}. However, despite its success, there is still room for improvement. The two main configurations of W7-X, the KJM (or so-called ``high mirror'') and the EIM (also known as ``standard'') are not optimized for simultaneously having low levels of radial and parallel neoclassical transport \cite{Beidler_2011, Beidler2021}: While W7-X EIM has small radial transport, it has intolerably large bootstrap current. Conversely, W7-X KJM displays small bootstrap current but larger levels of radial transport. Consequently, optimization of QI stellarators is a very active branch of research and, recently, much effort has been put in pushing forward the design and construction of quasi-isodynamic stellarators \cite{Sanchez_2023,velasco2023robust,RJorge_2022,camachomata_plunk_jorge_2022,goodman2022constructing}.

The QS subclass of omnigenous configurations is attractive as the neoclassical properties of such magnetic fields are isomorphic to those in a tokamak \cite{Pytte_Isomorphic,Boozer_Isomorphic}. Recently, it has been shown that it is possible to design QS magnetic fields with extremely low neoclassical losses \cite{Landreman_PreciseQS}. In contrast to QI configurations, QS stellarators are expected to have a substantial bootstrap current\footnote{With the exception of the quasi-poloidally symmetric magnetic field, which lies at the intersection of QI and QS configurations. However, quasi-poloidal symmetry is impossible to achieve near the magnetic axis, see e.g. \cite{Helander_2014}.} and its effect must be taken into account \cite{Landreman_SelfConsistent}. Examples of this subclass are the Helically Symmetric eXperiment (HSX) \cite{HSX} or the design of the National Compact Stellarator Experiment (NCSX) \cite{NCSX}. 

Typically, at each iteration of the optimization process a large number ($\sim$$10^2$) of magnetic configurations are generated. Therefore, in order to neoclassically optimize magnetic fields, it is required to be able to evaluate fast the neoclassical properties of each configuration. Due to this requirement, neoclassical properties are typically addressed indirectly. For instance, one can tailor the variation of the magnetic field strength $B$ on the flux-surface so that it nearly fulfils quasi-isodinamicity: the isolines of $B$ can be forced to close poloidally and the variance of the extrema of $B$ along field lines can be minimized.

A different approach relies on figures of merit, which are easy to calculate, for specific collisional regimes. For the $1/\nu$ regime, the code {\texttt{NEO}} \cite{Nemov1999EvaluationO1} computes the effective ripple $\epsilon_{\text{eff}}$, which encapsulates the dependence of radial neoclassical transport on the magnetic configuration. For transport within the flux surface, there exist long mean free path formulae for parallel flow and bootstrap current \cite{Shaing-Callen-1983,Nakajima-1988,helander_parra_newton_2017}. Although they can be computed very fast and capture some qualitative behaviour, these formulae are plagued with noise due to resonances in rational surfaces and, even with smoothing ad-hoc techniques, they are not accurate \cite{Landreman_SelfConsistent}. This lack of accuracy limits their application for optimization purposes. During the optimization process, an accurate calculation of the bootstrap current is required to account for its effect (e.g. for optimizing QS stellarators) or to keep it sufficiently small (when optimizing for quasi-isodinamicity). 

Recent developments allow direct optimization of radial neoclassical transport. Based on previous derivations \cite{Calvo_2017,dherbemont2022}, the code {\texttt{KNOSOS}} \cite{KNOSOSJCP,KNOSOSJPP} solves very fast an orbit-averaged drift-kinetic equation that is accurate for low collisionality regimes. {\texttt{KNOSOS}} is included in the stellarator optimization suite \texttt{STELLOPT} \cite{STELLOPT} and in the predictive transport frameworks {\texttt{TANGO}} \cite{Banon_Navarro_2023} and \texttt{TRINITY} \cite{Barnes_Trinity_2010}. However, the orbit-averaged equations solved by {\texttt{KNOSOS}} only describe radial transport at low collisionalities. 

In this work we present {\MONKES} (MONoenergetic Kinetic Equation Solver), a new neoclassical code conceived to satisfy the necessity of fast and accurate calculations of the bootstrap current for stellarator optimization. Specifically, {\MONKES} makes it possible to compute the monoenergetic coefficients $\widehat{D}_{ij}$ where $i,j\in\{1,2,3\}$ (their precise definition is given in section \ref{sec:DKE}). These nine coefficients encapsulate neoclassical transport across and within flux surfaces. The parallel flow of each species can be calculated in terms of the coefficients $\widehat{D}_{3j}$ \cite{Taguchi,Sugama-PENTA,Sugama2008,MaasbergMomentumCorrection}. In the absence of externally applied loop voltage, the bootstrap current is driven by the radial electric field and gradients of density and temperature. The so-called bootstrap current coefficient $\widehat{D}_{31}$ is the one that relates the parallel flow to these gradients. The six remaining coefficients $\widehat{D}_{ij}$ for $i\in\{1,2\}$ allow to compute the flux of particles and heat across the flux surface. 

{\MONKES} also computes fast the radial transport coefficients. Although at low collisionality it is not as fast as the orbit-averaged code \texttt{KNOSOS}, {\MONKES} can compute the transition from the $1/\nu$ and $\sqrt{\nu}$-$\nu$ regimes to the plateau regime or the banana regime. The plateau regime may be relevant close to the edge, while the banana regime may be necessary for stellarators very close to perfect omnigeneity. Apart from optimization, {\MONKES} can find many other applications. For instance, it can be used for the analysis of experimental discharges or also be included in predictive transport frameworks.

This paper is organized as follows: in section \ref{sec:DKE}, we introduce the drift-kinetic equation solved by {\MONKES} and the transport coefficients that it computes. In section \ref{sec:Algorithm}, we explain the algorithm used to solve the drift-kinetic equation and its implementation. In section \ref{sec:Results_Benchmark}, by means of a convergence study, we demonstrate that {\MONKES} can be used to compute accurate monoenergetic coefficients at low collisionality very fast for the $1/\nu$ and $\sqrt{\nu}$-$\nu$ regimes \cite{dherbemont2022}. In order to show this, {\MONKES} results are also benchmarked against {\DKES} \cite{DKES1986,VanRij_1989} and, when necessary, against {\texttt{SFINCS}} \cite{Landreman_2014}. Finally, in section \ref{sec:Conclusions} we summarize the results and discuss future lines of work.

\section{Drift-kinetic equation and transport coefficients}
\label{sec:DKE}
{\MONKES} solves the drift-kinetic equation 
\begin{align}
	(v \xi \vb*{b}  + \vb*{v}_E) \cdot \nabla h_a 
	+
	v\nabla \cdot \vb*{b} \frac{(1-\xi^2)}{2}  \pdv{h_a}{\xi}  
	& - \nu^{a} \Lorentz h_a
	\nonumber \\
	& = S_a,
	\label{eq:DKE_Original}
\end{align}
where $\vb*{b}:= \vb*{B}/B$ is the unit vector tangent to magnetic field lines and we have employed as velocity coordinates the cosine of the pitch-angle $\xi := \vb*{v}\cdot\vb*{b}/|\vb*{v}|$ and the magnitude of the velocity $v:=|\vb*{v}|$. 

We assume that the magnetic configuration has nested flux-surfaces. We denote by $\psi\in[0,\psi_{\text{lcfs}}]$ a radial coordinate that labels flux-surfaces, where $\psi_{\text{lcfs}}$ denotes the label of the last closed flux-surface. In equation (\ref{eq:DKE_Original}), $h_a$ is the non-adiabatic component of the deviation of the distribution function from a local Maxwellian for a plasma species $a$ 
\begin{align}
	f_{\text{M}a}(\psi, v) :=   n_a(\psi)  \pi^{-3/2}  {v_{\text{t}a}^{-3}(\psi)}  \exp(-\frac{v^2}{v_{\text{t}a}^2(\psi)}).
\end{align}
Here, $n_a$ is the density of species $a$, $v_{\text{t}a} := \sqrt{2T_a/m_a}$ is its thermal velocity, $T_a$ its temperature (in energy units) and $m_a$ its mass. 

%

For the convective term in equation (\ref{eq:DKE_Original})
\begin{align}
	\vb*{v}_E 
	:= 
	\frac{\vb*{E}_0\times\vb*{B}}{\mean*{B^2}} 
	= 
	- 
	\frac{E_\psi}{\mean*{B^2}}\vb*{B}\times \nabla\psi
	\label{eq:Incompressible_ExB_definition}
\end{align}
denotes the incompressible $\vb*{E}\times\vb*{B}$ drift approximation \cite{dherbemont2022} and $\vb*{E}_0 = E_\psi(\psi) \nabla \psi$ is the electrostatic piece of the electric field $\vb*{E}$ perpendicular to the flux-surface. The symbol $\mean*{...}$ stands for the flux-surface average operation. Denoting by $V(\psi)$ the volume enclosed by the flux-surface labelled by $\psi$, the flux-surface average of a function $f$ can be defined as the limit
\begin{align}
	\mean*{f}
	:=
	\lim_{\delta \psi \rightarrow 0} 
	\dfrac{\int_{V(\psi+\delta\psi)} f \dd[3]{\vb*{r}}- \int_{V(\psi)} f\dd[3]{\vb*{r}}}
	{V(\psi+\delta\psi) - V(\psi)},
	\label{eq:FSA}
\end{align}
where $\dd[3]{\vb*{r}}$ is the spatial volume form.

We denote the Lorentz pitch-angle scattering operator by $\Lorentz$, which in coordinates $(\xi,v)$ takes the form
\begin{align}
	\Lorentz   := \frac{1}{2}  \pdv{\xi}\left( (1-\xi^2)\pdv{}{\xi} \right).
	\label{eq:Pitch_angle_scattering_operator}
\end{align}
In the collision operator, $\nu^a(v) =\sum_{b}\nu^{ab}(v)$ and
\begin{align}
	\nu^{ab}(v) := 
	\frac{4 \pi n_b e_a^2 e_b^2}
	{m_a^2 v_{\text{t}a}^3}
	\log\Lambda
	\frac{ \erf(v/v_{\text{t}b}) - G(v/v_{\text{t}b})}{v^3/v_{\text{t}a}^3}
\end{align}
stands for the pitch-angle collision frequency between species $a$ and $b$. We denote the respective charges of each species by $e_a$ and $e_b$, the Chandrasekhar function by $G(x)=\left[\erf(x) - (2x/\sqrt{\pi}) \exp(-x^2)\right]/(2x^2)$, $\erf(x)$ is the error function and $\log\Lambda$ is the Coulomb logarithm \cite{Helander_2005}.

On the right-hand-side of equation (\ref{eq:DKE_Original}) 
\begin{align}
	S_a 
	& :=  
	- \vb*{v}_{\text{m} a} \cdot \nabla \psi 
	\left(
	A_{1a} 
	+  \frac{v^2}{v_{\text{t}a}^2}
	A_{2a}
	\right)
	f_{\text{M}a}
	\nonumber \\ 
	& + 
	B v \xi A_{3a}f_{\text{M}a}
	\label{eq:DKE_Original_Source}
\end{align}
is the source term, 
\begin{align}
	\vb*{v}_{\text{m} a}\cdot\nabla\psi
	=
	-\frac{Bv^2}{\Omega_a}
	\frac{1+\xi^2}{2B^3}
	\vb*{B}\times\nabla\psi \cdot \nabla B 
\end{align}
is the expression of the radial magnetic drift assuming ideal magnetohydrodynamical equilibrium, $\Omega_a = e_a B / m_a$ is the gyrofrecuency of species $a$ and the flux-functions 
\begin{align}
	A_{1a}(\psi) & := \dv{\ln n_a}{\psi} - \frac{3}{2} \dv{\ln T_a}{\psi} - \frac{e_a E_\psi}{T_a}, 
	\\
	A_{2a}(\psi) & := \dv{\ln T_a}{\psi} , 
	\\
	A_{3a}(\psi) & :=  \frac{e_a }{T_a} \frac{\mean*{\vb*{E}\cdot\vb*{B}}}{\mean*{B^2}}
\end{align}
are the so-called thermodynamical forces.

Mathematically speaking, there are still two additional conditions to completely determine the solution to equation (\ref{eq:DKE_Original}). First, equation (\ref{eq:DKE_Original}) must be solved imposing regularity conditions at $\xi =\pm 1$
\begin{align}
	\eval{\left((1-\xi^2) \pdv{h_a}{\xi}\right)}_{\xi =\pm 1} = 0.
	\label{eq:Regularity_conditions}
\end{align}
Second, as the differential operator on the left-hand-side of equation (\ref{eq:DKE_Original}) has a non trivial kernel, the solution to equation (\ref{eq:DKE_Original}) is determined up to an additive function $g(\psi,v)$. This function is unimportant as it does not contribute to the neoclassical transport quantities of interest. Nevertheless, in order to have a unique solution to the drift-kinetic equation, it must be fixed by imposing an appropriate additional constraint. We will select this free function (for fixed $(\psi,v)$) by imposing
\begin{align}
	\mean*{  \int_{-1}^{1} h_a \dd{\xi}  } = C,
	\label{eq:kernel_elimination_condition}
\end{align}
for some $C\in\mathbb{R}$. We will discuss this further in section \ref{sec:Algorithm}. 

The drift-kinetic equation (\ref{eq:DKE_Original}) is the one solved by the standard neoclassical code {\DKES} \cite{DKES1986, VanRij_1989} using a variational principle. 
Although the main feature of the code \texttt{SFINCS} \cite{Landreman_2014} is to solve a more complete neoclassical drift-kinetic equation, it can also solve equation (\ref{eq:DKE_Original}).

Taking the moments $\{\vb*{v}_{\text{m} a} \cdot \nabla\psi,  (v^2/v_{\text{t}a}^2)\vb*{v}_{\text{m} a} \cdot \nabla\psi, v\xi B/B_0\}$ of $h_a$ and then the flux-surface average yields, respectively, the radial particle flux, the radial heat flux and the parallel flow
\begin{align}
	\mean*{\vb*{\Gamma}_a \cdot \nabla \psi} & := 
	\mean*{
		\int
		\vb*{v}_{\text{m} a} \cdot \nabla\psi	
		\ h_a
		\dd[3]{\vb*{v}}
	},
	\label{eq:Particle_flux_Original}
	\\
	\mean*{\frac{\vb*{Q}_a \cdot \nabla \psi}{T_a}} & := 
	\mean*{
		\int
		\frac{v^2}{v_{\text{t}a}^2}\vb*{v}_{\text{m} a} \cdot \nabla\psi	
		\ h_a
		\dd[3]{\vb*{v}}
	},
	\label{eq:Heat_flux_Original}
	\\
	\frac{\mean*{n_a \vb*{V}_{a} \cdot\vb*{B}}}{B_{0}} & :=
	\mean*{
		\frac{B}{B_0}
		\int
		v \xi 
		\ h_a
		\dd[3]{\vb*{v}}
	},
	\label{eq:Parallel_flow_Original}
\end{align}
where $B_0(\psi)$ is a reference value for the magnetic field strength on the flux-surface (its explicit definition is given in section \ref{sec:Algorithm}).

It is a common practice for linear drift-kinetic equations (e.g. \cite{DKES1986, Beidler_2011,Landreman_2014}) to apply superposition and split $h_a$ into several additive terms. As in the drift-kinetic equation (\ref{eq:DKE_Original}) there are no derivatives or integrals along $\psi$ nor $v$, it is convenient to use the splitting
\begin{align}
	h_a 
	= 
	f_{\text{M}a}
	\left[
	\frac{B v}{\Omega_a} 
	\left(
	A_{1a} f_1 
	+ 
	A_{2a}  
	\frac{v^2}{v_{\text{t}a}^2}f_2
	\right)
	+
	B_0 A_{3a} f_3
	\right].
	\label{eq:Distribution_function_superposition}
\end{align}
The splitting is chosen so that the functions $\{f_j\}_{j=1}^{3}$ are solutions to
\begin{align}
	\xi \vb*{b}  \cdot 
	\nabla f_j
	& +
	\nabla \cdot \vb*{b} \frac{(1-\xi^2)}{2}  \pdv{f_j}{\xi}  
	\nonumber\\
	&
	- 
	\frac{\widehat{E}_\psi}{\mean*{B^2}}
	\vb*{B}\times \nabla\psi\cdot \nabla f_j
	\label{eq:DKE}
	- \hat{\nu}\Lorentz f_j
	=  s_j, \quad 
\end{align}
for $j=1,2,3$, where $\hat{\nu} := \nu(v) / v$ and $\widehat{E}_\psi := {E}_\psi/v$. The source terms are defined as
\begin{align}
	s_1 := - \vb*{v}_{\text{m} a} \cdot \nabla\psi \frac{\Omega_a}{B v^2},
	\quad
	s_2 :=  s_1, 
	\quad
	s_3 := \xi \frac{B}{B_0}.
	\label{eq:DKE_Sources}
\end{align} 
Note that each source $s_j$ corresponds to one of the three thermodynamic forces on the right-hand side of definition (\ref{eq:DKE_Original_Source}).

The relation between $h_a$ and $f_j$ given by equation (\ref{eq:Distribution_function_superposition}) is such that the transport quantities (\ref{eq:Particle_flux_Original}), (\ref{eq:Heat_flux_Original}) and (\ref{eq:Parallel_flow_Original}) can be written in terms of four transport coefficients which, for fixed $(\hat{\nu}, \widehat{E}_\psi)$, depend only on the magnetic configuration. As $\dv*{\hat{\nu}}{v}$ never vanishes, the dependence of $f_j$ on the velocity $v$ can be parametrized by its dependence on $\hat{\nu}$. Thus, for fixed $(\hat{\nu}, \widehat{E}_\psi)$, equation (\ref{eq:DKE}) is completely determined by the magnetic configuration. Hence, its unique solutions $f_j$ that satisfy conditions (\ref{eq:Regularity_conditions}) and (\ref{eq:kernel_elimination_condition}) are also completely determined by the magnetic configuration. The assumptions that lead to $\psi$ and $v$ appearing as parameters in the drift-kinetic equation (\ref{eq:DKE_Original}) comprise the so-called local monoenergetic approximation to neoclassical transport (see e.g. \cite{Landreman_Monoenergetic}).

Using splitting (\ref{eq:Distribution_function_superposition}) we can write the transport quantities (\ref{eq:Particle_flux_Original}), (\ref{eq:Heat_flux_Original}) and (\ref{eq:Parallel_flow_Original}) in terms of the Onsager matrix
\begin{align}
	&\Matrix{c}
	{
		\mean*{\vb*{\Gamma}_a \cdot \nabla \psi} \\
		\mean*{ \dfrac{\vb*{Q}_a \cdot \nabla \psi}{T_a} }     \\ 
		\dfrac{\mean*{n_a \vb*{V}_{a} \cdot\vb*{B}}}{B_0}
	}
	=
	\Matrix{ccc}
	{
		L_{11a} & L_{12a}  & L_{13a} \\
		L_{21a} & L_{22a}  & L_{23a} \\
		L_{31a} & L_{32a}  & L_{33a} 
	}
	\Matrix{c}
	{ 
		A_{1a} \\
		A_{2a} \\
		A_{3a} 
	}.
\end{align}
Here, we have defined the thermal transport coefficients as 
\begin{align}
	L_{ija} :=    
	\int_{0}^{\infty}
	2\pi v^2
	f_{\text{M}a} 
	w_i w_j 
	D_{ija} 
	\dd{v}, \ \ %
\end{align}
where $w_1=w_3=1$, $w_2=v^2/v_{\text{t}a}^2$ and we have used that $\int g\dd[3]{\vb*{v}} = 2\pi \int_{0}^{\infty}\int_{-1}^{1} g v^2 \dd{\xi}\dd{v}$ for any integrable function $g(\xi,v)$. The quantities $D_{ija}$ are defined as
\begin{align}
	D_{ija} & :=-\frac{B^2v^3}{\Omega_a^2} \widehat{D}_{ij}, &\quad i,j \in\{1,2\},
	\\
	D_{i3a} & :=  
	- \frac{B_0 B v^2}{\Omega_a} \widehat{D}_{i3}, &\quad i \in\{1,2\},
	\\
	D_{3ja} & := \frac{B v^2}{\Omega_a} \widehat{D}_{3j}, &\quad j \in\{1,2\},
	\\
	D_{33a} & := v B_0 \widehat{D}_{33}, &
\end{align}
where  
\begin{align}
	\widehat{D}_{ij}(\psi,v) := \mean*{ \int_{-1}^{1}  s_i f_j   \dd{\xi} }, \quad i,j\in\{1,2,3\}
	\label{eq:Monoenergetic_geometric_coefficients}
\end{align} 
are the monoenergetic geometric coefficients. Note that (unlike $D_{ija}$) the monoenergetic geometric coefficients $\widehat{D}_{ij}$ do not depend on the species for fixed $\hat{\nu}$ (however the correspondent value of $v$ associated to each $\hat{\nu}$ varies between species) and depend only on the magnetic geometry. In general, four independent monoenergetic geometric coefficients can be obtained by solving (\ref{eq:DKE}): $\widehat{D}_{11}$, $\widehat{D}_{13}$, $\widehat{D}_{31}$ and $\widehat{D}_{33}$. However, when the magnetic field possesses stellarator symmetry \cite{DEWAR1998275} or there is no radial electric field, Onsager symmetry implies $\widehat{D}_{13} = -\widehat{D}_{31}$ \cite{VanRij_1989} making only three of them independent (for further details see \ref{sec:Appendix_Onsager_symmetry}). Hence, obtaining the transport coefficients for all species requires to solve (\ref{eq:DKE}) for two different source terms $s_1$ and $s_3$. The algorithm for solving equation (\ref{eq:DKE}) is described in section \ref{sec:Algorithm}. 

Finally, we briefly comment on the validity of the coefficients provided by equation (\ref{eq:DKE}) for the calculation of the bootstrap current. The pitch-angle scattering collision operator used in equation (\ref{eq:DKE_Original}) lacks parallel momentum conservation. Besides, the pitch-angle scattering operator is not adequate for calculating parallel flow of electrons, which is a quantity required to compute the bootstrap current. Hence, in principle, the parallel transport directly predicted by equation (\ref{eq:DKE_Original}) is not correct. Fortunately, there exist techniques  \cite{Taguchi,Sugama-PENTA,Sugama2008,MaasbergMomentumCorrection} to calculate the radial and parallel transport associated to more accurate momentum conserving collision operators by just solving the simplified drift-kinetic equation (\ref{eq:DKE}). This has been done successfully in the past by the code \texttt{PENTA} \cite{Sugama-PENTA, Spong-PENTA}, using the results of {\DKES}. Nevertheless, the momentum restoring technique is not needed for minimizing the bootstrap current. In the method presented in section V of \cite{MaasbergMomentumCorrection}, when there is no net parallel inductive electric field (i.e. $A_{3a}=0$), the parallel flow with the correct collision operator for any species vanishes when two integrals in $v$ of $\widehat{D}_{31}$ vanish. Thus, minimizing $\widehat{D}_{31}$ translates in a minimization of the parallel flows of all species involved in the bootstrap current calculation, and therefore of this current.  

\section{Numerical method}
\label{sec:Algorithm}
 In this section we describe the algorithm to numerically solve the drift-kinetic equation (\ref{eq:DKE}) and its implementation. The algorithm, based on the tridiagonal representation of the drift-kinetic equation, emerges naturally when the velocity coordinate $\xi$ is discretized using a Legendre spectral method.

First, in subsection \ref{subsec:Legendre_expansion} we will present the algorithm in a formal way. We will use (right-handed) Boozer coordinates\footnote{Even though we use Boozer coordinates, we want to stress out that the algorithm presented in subsection \ref{subsec:Legendre_expansion} is valid for any set of spatial coordinates in which $\psi$ labels flux-surfaces and the two remaining coordinates parametrize the flux-surface.} $(\psi,\theta,\zeta)\in[0,\psi_{\text{lcfs}}]\times[0,2\pi)\times[0,2\pi/N_p)$. The integer $N_p\ge 1$ denotes the number of toroidal periods of the device. The radial coordinate is selected so that $2\pi \psi$ is the toroidal flux of the magnetic field and $\theta$, $\zeta$ are respectively the poloidal and toroidal (in a single period) angles. In these coordinates, the magnetic field can be written as
\begin{align}
	\vb*{B} & = \nabla\psi \times \nabla\theta - \iota(\psi) \nabla\psi \times \nabla\zeta 
	\nonumber\\
	& = B_\psi(\psi,\theta,\zeta) \nabla \psi + B_\theta(\psi) \nabla \theta + B_\zeta(\psi) \nabla \zeta,
	\label{eq:Magnetic_field_Boozer}
\end{align}
and the Jacobian of the transformation reads 
\begin{align}
	\sqrt{g}(\psi,\theta,\zeta) 
	:=( 
	\nabla\psi \times \nabla \theta \cdot \nabla\zeta  
	)^{-1} 
	= 
	\frac{B_\zeta + \iota B_\theta}{B^2},
	\label{eq:Jacobian_Boozer}
\end{align} 
where $\iota :=\vb*{B} \cdot \nabla \theta / \vb*{B} \cdot \nabla \zeta $ is the rotational transform. The flux-surface average operation (\ref{eq:FSA}) is written in Boozer angles as
\begin{align}
	\mean*{f}
	=
	\left(\dv{V}{\psi}\right)^{-1}
	\oint\oint
	f
	\sqrt{g}
	\dd{\theta}\dd{\zeta}
	.
	\label{eq:FSA_Boozer}
\end{align}

We define the reference value for the magnetic field strength $B_0$ introduced in definition (\ref{eq:Parallel_flow_Original}) as the $(0,0)$ Fourier mode of the magnetic field strength. Namely, 
\begin{align}
	B_0(\psi) := \frac{N_p}{4\pi^2 } 
	\oint\oint
	B(\psi,\theta,\zeta)
	\dd{\theta}\dd{\zeta}.
\end{align}

Using (\ref{eq:Magnetic_field_Boozer}) and (\ref{eq:Jacobian_Boozer}), the spatial differential operators present in the drift-kinetic equation (\ref{eq:DKE}) can be expressed in these coordinates as
\begin{align}
	\vb*{b} \cdot \nabla & = 
	\frac{B}{B_\zeta + \iota B_\theta}
	\left(
	\iota \pdv{\theta}
	+ 
	\pdv{\zeta} 
	\right), \label{eq:Parallel_streaming_spatial_operator}
	\\
	\vb*{B}\times\nabla\psi \cdot \nabla & = 
	\frac{B^2}{B_\zeta + \iota B_\theta}
	\left(
	B_\zeta \pdv{\theta}
	-
	B_\theta \pdv{\zeta}
	\right). \label{eq:ExB_spatial_operator}
\end{align}

After the explanation of the algorithm, in subsection \ref{subsec:Algorithm_Implementation} its implementation in {\MONKES} is described. In order to ease the notation, in subsections \ref{subsec:Legendre_expansion} and \ref{subsec:Algorithm_Implementation} we drop when possible the subscript $j$ that labels every different source term.  Also, as $\psi$ and $v$ act as mere parameters, we will omit their dependence and functions of these two variables will be referred to as constants. 

\subsection{Legendre polynomial expansion}\label{subsec:Legendre_expansion}
The algorithm is based on the approximate representation of the distribution function $f$ by a truncated Legendre series. We will search for approximate solutions to equation (\ref{eq:DKE}) of the form
\begin{align}
	f(\theta,\zeta,\xi) = \sum_{k=0}^{N_\xi} f^{(k)}(\theta,\zeta) P_k(\xi), \label{eq:Legendre_expansion}
\end{align} 
where $f^{(k)} = \mean*{f,P_k}_\Lorentz / \mean*{P_k,P_k}_\Lorentz$ is the $k-$th Legendre mode of $f(\theta,\zeta,\xi)$ (see \ref{sec:Appendix_Legendre}) and $N_\xi$ is an integer greater or equal to 1. As mentioned in \ref{sec:Appendix_Legendre}, the expansion in Legendre polynomials (\ref{eq:Legendre_expansion}) ensures that the regularity conditions (\ref{eq:Regularity_conditions}) are satisfied. Of course, in general, the exact solution to equation (\ref{eq:DKE}) does not have a finite Legendre spectrum, but taking $N_\xi$ sufficiently high in expansion (\ref{eq:Legendre_expansion}) yields an approximate solution to the desired degree of accuracy (in infinite precision arithmetic).

In \ref{sec:Appendix_Legendre} we derive explicitly the projection of each term of the drift-kinetic equation (\ref{eq:DKE}) onto the Legendre basis when the representation (\ref{eq:Legendre_expansion}) is used. When doing so, we obtain that the Legendre modes of the drift-kinetic equation have the tridiagonal representation  
\begin{align}
	L_k f^{(k-1)} + D_k f^{(k)} + U_k f^{(k+1)} = s^{(k)},  
	\label{eq:DKE_Legendre_expansion}
\end{align}
for $k=0,1,\ldots ,N_\xi$, where we have defined for convenience $f^{(-1)}:=0$ and from expansion (\ref{eq:Legendre_expansion}) it is clear that $f^{(N_\xi+1)}=0$. Analogously to (\ref{eq:Legendre_expansion}) the source term is expanded as $s=\sum_{k=0}^{N_\xi} s^{(k)} P_k$. For the sources given by (\ref{eq:DKE_Sources}) this expansion is exact when $N_\xi\ge2$ as $s_j^{(k)}=0$ for $k\ge 3$. The spatial differential operators read 
\begin{align}
	L_k & = 
	\frac{k}{2k-1} 
	\left(
	\vb*{b} \cdot \nabla 
	+
	\frac{k-1}{2}
	\vb*{b}\cdot\nabla \ln B
	\right), \label{eq:DKE_Legendre_expansion_Lower}
	\\ 
	D_k & = - 
	\frac{\widehat{E}_\psi}{\mean*{B^2}}
	\vb*{B}\times \nabla\psi  \cdot \nabla 
	+  
	\frac{k(k+1)}{2}
	\hat{\nu} , \label{eq:DKE_Legendre_expansion_Diagonal}
	\\
	U_k & =  
	\frac{k+1}{2k+3} 
	\left(
	\vb*{b} \cdot \nabla 
	-
	\frac{k+2}{2}
	\vb*{b}\cdot\nabla \ln B
	\right). \label{eq:DKE_Legendre_expansion_Upper}
\end{align}
Thanks to its tridiagonal structure, the system of equations (\ref{eq:DKE_Legendre_expansion}) can be inverted using the standard Gaussian elimination algorithm for block tridiagonal matrices.

Before introducing the algorithm we will explain how to fix the free constant of the solution to equation (\ref{eq:DKE_Legendre_expansion}) so that it can be inverted. Note that the aforementioned kernel of the drift-kinetic equation translates in the fact that $f^{(0)}$ is not completely determined from equation (\ref{eq:DKE_Legendre_expansion}). To prove this, we inspect the modes $k=0$ and $k=1$ of equation (\ref{eq:DKE_Legendre_expansion}), which are the ones that involve $f^{(0)}$. From expression (\ref{eq:ExB_spatial_operator}) we can deduce that the term $D_0 f^{(0)} + U_0 f^{(1)} $ is invariant if we add to $f^{(0)}$ any function of $B_\theta \theta + B_\zeta  \zeta$. For $\widehat{E}_\psi\ne 0$, functions of $B_\theta \theta + B_\zeta  \zeta$ lie on the kernel of $\vb*{B}\times\nabla \psi \cdot \nabla$ and for $\widehat{E}_\psi = 0$, $D_0$ is identically zero. Besides, the term $L_1 f^{(0)} + D_1 f^{(1)} + U_1 f^{(2)}$ remains invariant if we add to $f^{(0)} $ any function of $\theta-\iota\zeta$ (the kernel of $L_1=\vb*{b}\cdot\nabla$ consists of these functions). For ergodic flux-surfaces, the only continuous functions on the torus that belong to the kernel of $L_1$ are constants. Thus, equation (\ref{eq:DKE_Legendre_expansion}) is unaltered when we add to $f^{(0)}$ any constant (a function that belongs simultaneously to the kernels of $\vb*{B}\times\nabla \psi \cdot \nabla$ and $\vb*{b}\cdot\nabla$). A constraint equivalent to condition (\ref{eq:kernel_elimination_condition}) is to fix the value of the $0-$th Legendre mode of the distribution function at a single point of the flux-surface. For example,
\begin{align}
	f^{(0)}(0,0)=0, \label{eq:kernel_elimination_condition_Legendre}
\end{align}
which implicitly fixes the value of the constant $C$ in (\ref{eq:kernel_elimination_condition}).
With this condition, equation (\ref{eq:DKE_Legendre_expansion}) has a unique solution and its left-hand-side can be inverted  to solve for $f^{(k)}$ in two scenarios: when the flux-surface is ergodic and in rational surfaces when $\widehat{E}_\psi\ne0$ (further details on its invertibility are given in \ref{sec:Appendix_Invertibility}). Note that, as expansion (\ref{eq:Legendre_expansion}) is finite and representation (\ref{eq:DKE_Legendre_expansion}) is non diagonal, the functions $f^{(k)}$ obtained from inverting (\ref{eq:DKE_Legendre_expansion}) constrained by (\ref{eq:kernel_elimination_condition_Legendre}) are approximations to the first $N_\xi+1$ Legendre modes of the exact solution to (\ref{eq:DKE}) satisfying (\ref{eq:kernel_elimination_condition}) (further details at the end of \ref{sec:Appendix_Legendre}).

The algorithm for solving the truncated drift-kinetic equation (\ref{eq:DKE_Legendre_expansion}) consists of two steps. 
\begin{enumerate}
	\item \textbf{Forward elimination}
\end{enumerate} 	
Starting from $\Delta_{N_\xi} = D_{N_\xi}$ and $\sigma^{(N_\xi)} = s^{(N_\xi)}$ we can obtain recursively the operators
\begin{align}
	\Delta_k = D_k - U_{k} \Delta_{k+1}^{-1} L_{k+1}, 
	\label{eq:Schur_complements}
\end{align} 
and the sources
\begin{align}
	\sigma^{(k)} = s^{(k)} - U_k \Delta_{k+1}^{-1}    \sigma^{(k+1)},
	\label{eq:Forward_elimination_sources}
\end{align}
for $k=N_\xi-1, N_\xi-2, \ldots, 0$ (in this order). Equations (\ref{eq:Schur_complements}) and (\ref{eq:Forward_elimination_sources}) define the forward elimination. With this procedure we can transform equation (\ref{eq:DKE_Legendre_expansion}) to the equivalent system
\begin{align}
	L_{k} f^{(k-1)} + \Delta_{k} f^{(k)} = \sigma^{(k)},
	\label{eq:DKE_Forward_elimination}
\end{align}
for $k=0,1, \ldots, N_\xi$. Note that this process corresponds to perform formal Gaussian elimination over 
\begin{align}
	\Matrix{ccc|c}
	{
		L_{k} & D_{k} & U_k  & s^{(k)} \\
		0& L_{k+1} & \Delta_{k+1}  & \sigma^{(k+1)} \\
	}
	,
	\label{eq:DKE_Forward_Elimination}
\end{align}
to eliminate $U_k$ in the first row.

\begin{enumerate}[resume]
	\item \textbf{Backward substitution}
\end{enumerate}
Once we have the system of equations in the form (\ref{eq:DKE_Forward_elimination}) it is immediate to solve recursively
\begin{align}
	f^{(k)} = 
	\Delta_k^{-1}
	\left( 
	\sigma^{(k)} -  L_{k} f^{(k-1)} 
	\right), 	\label{eq:DKE_Backward_Substitution}
\end{align}
for $k=0,1,...,N_\xi$ (in this order). Here, $\Delta_0^{-1} \sigma^{(0)}$ denotes the unique solution to $\Delta_0 f^{(0)} = \sigma^{(0)} $ that satisfies (\ref{eq:kernel_elimination_condition_Legendre}). As $L_1= \vb*{b}\cdot \nabla$, using expression (\ref{eq:Parallel_streaming_spatial_operator}), it is clear from equation (\ref{eq:DKE_Backward_Substitution}) that the integration constant does not affect the value of $f^{(1)}$.

We can apply this algorithm to solve equation (\ref{eq:DKE}) for $f_1$, $f_2$ and $f_3$ in order to compute approximations to the transport coefficients. In terms of the Legendre modes of $f_1$, $f_2$ and $f_3$, the monoenergetic geometric coefficients from definition (\ref{eq:Monoenergetic_geometric_coefficients}) read
\begin{align}
	\widehat{D}_{11} & = 2\mean*{s_1^{(0)} f_1^{(0)}} + \frac{2}{5}\mean*{s_1^{(2)} f_1^{(2)}}, 
	\label{eq:Gamma_11_Legendre}\\ 
	\widehat{D}_{31} & = \frac{2}{3} \mean*{\frac{B}{B_0} f_1^{(1)}},\label{eq:Gamma_31_Legendre}\\ 
	\widehat{D}_{13} & = 2\mean*{s_1^{(0)} f_3^{(0)}} + \frac{2}{5}\mean*{s_1^{(2)} f_3^{(2)}}, \label{eq:Gamma_13_Legendre}\\ 
	\widehat{D}_{33} & =\frac{2}{3} \mean*{\frac{B}{B_0} f_3^{(1)}}, \label{eq:Gamma_33_Legendre}
\end{align}
where $3s_1^{(0)} /2= 3s_1^{(2)} = \vb*{B}\times\nabla\psi \cdot \nabla B / B^3$. Note  from expressions (\ref{eq:Gamma_11_Legendre}), (\ref{eq:Gamma_31_Legendre}), (\ref{eq:Gamma_13_Legendre}) and (\ref{eq:Gamma_33_Legendre}) that, in order to compute the monoenergetic geometric coefficients $\widehat{D}_{ij}$, we only need to calculate the Legendre modes $k=0,1,2$ of the solution and we can stop the backward substitution (\ref{eq:DKE_Backward_Substitution}) at $k=2$. In the next subsection we will explain how {\MONKES} solves equation (\ref{eq:DKE_Legendre_expansion}) using this algorithm.

\subsection{Spatial discretization and algorithm implementation}\label{subsec:Algorithm_Implementation}
The algorithm described above allows, in principle, to compute the exact solution to the truncated drift-kinetic equation (\ref{eq:DKE_Legendre_expansion}) which is an approximate solution to (\ref{eq:DKE}). However, to our knowledge, it is not possible to give an exact expression for the operator $\Delta_k^{-1}$ except for $k=N_\xi \ge 1$. Instead, we are forced to compute an approximate solution to (\ref{eq:DKE_Legendre_expansion}).
In order to obtain an approximate solution of equation (\ref{eq:DKE_Legendre_expansion}) we assume that each $f^{(k)}$ has a finite Fourier spectrum so that it can be expressed as
\begin{align}	
	f^{(k)}(\theta,\zeta)
	& 
	=
	\vb*{I}(\theta,\zeta)
	\cdot
	\vb*{f}^{(k)},
	\label{eq:Fourier_expansion_f_k}
\end{align}
where the Fourier interpolant row vector map $\vb*{I}(\theta,\zeta)$ is defined at \ref{sec:Appendix_Fourier} and the column vector $\vb*{f}^{(k)}\in\mathbb{R}^{N_{\text{fs}}}$
contains $f^{(k)}$ evaluated at the equispaced grid points
\begin{align}
	\theta_i & = 2\pi i / N_\theta, \quad & i=0,1,\ldots, N_\theta-1,  \label{eq:Theta_grid}
	\\ 
	\zeta_j & = 2\pi j / (N_\zeta N_p), \quad & j=0,1,\ldots, N_\zeta-1. \label{eq:Zeta_grid}
\end{align}
Here, $N_{\text{fs}}:=N_\theta N_\zeta$ is the number of points in which we discretize the flux-surface being $N_\theta$ and $N_\zeta$ respectively the number of points in which we divide the domains of $\theta$ and $\zeta$. In general, the solution to equation (\ref{eq:DKE_Legendre_expansion}) has an infinite Fourier spectrum and cannot exactly be written as (\ref{eq:Fourier_expansion_f_k}) but, taking sufficiently large values of $N_\theta$ and $N_\zeta$, we can approximate the solution to equation (\ref{eq:DKE_Legendre_expansion}) to arbitrary degree of accuracy (in infinite precision arithmetic). As explained in \ref{sec:Appendix_Fourier}, introducing the Fourier interpolant (\ref{eq:Fourier_expansion_f_k}) in equation (\ref{eq:DKE_Legendre_expansion}) and then evaluating the result at the grid points provides a system of $N_{\text{fs}}\times(N_\xi+1)$ equations which can be solved for $\{\vb*{f}^{(k)}\}_{k=0}^{N_\xi}$. This system of equations is obtained by substituting the operators $L_k$, $D_k$, $U_k$ in equation (\ref{eq:DKE_Legendre_expansion}) by the $N_{\text{fs}}\times N_{\text{fs}}$ matrices $\vb*{L}_k$, $\vb*{D}_k$, $\vb*{U}_k$, defined in \ref{sec:Appendix_Fourier}. Thus, we discretize (\ref{eq:DKE_Legendre_expansion}) as %
\begin{align}
	\vb*{L}_k  \vb*{f}^{(k-1)} + \vb*{D}_k  \vb*{f}^{(k)} + \vb*{U}_k   \vb*{f}^{(k+1)} = \vb*{s}^{(k)},   \label{eq:DKE_Legendre_expansion_Fourier_collocation}
\end{align}
for $k=0,1\ldots, N_\xi$ where $\vb*{s}^{(k)}\in\mathbb{R}^{N_{\text{fs}}}$
contains $s^{(k)}$ evaluated at the equispaced grid points. This system has a block tridiagonal structure and the algorithm presented in subsection \ref{subsec:Legendre_expansion} can be applied. We just have to replace in equations (\ref{eq:Schur_complements}), (\ref{eq:Forward_elimination_sources}) and (\ref{eq:DKE_Backward_Substitution}) the operators and functions by their respective matrix and vector analogues, which we denote by boldface letters. 

The matrix approximation to the forward elimination procedure given by equations (\ref{eq:Schur_complements}) and (\ref{eq:Forward_elimination_sources}) reads
\begin{align}
	\vb*{\Delta}_k & = \vb*{D}_k - \vb*{U}_{k} \vb*{\Delta}_{k+1}^{-1} \vb*{L}_{k+1}, 
	\label{eq:Schur_complements_matrix}
	\\
	\vb*{\sigma}^{(k)} & = \vb*{s}^{(k)} - \vb*{U}_{k}  \vb*{\Delta}_{k+1}^{-1}    \vb*{\sigma}^{(k+1)},
	\label{eq:Forward_elimination_sources_matrix}
\end{align}
for $k=N_\xi-1, N_\xi-2, \ldots, 0$ (in this order). Thus, starting from $\vb*{\Delta}_{N_\xi}=\vb*{D}_{N_\xi}$ and $\vb*{\sigma}^{(N_\xi)}=\vb*{s}^{(N_\xi)}$ all the matrices $\vb*{\Delta}_k$ and the vectors $\vb*{\sigma}^{(k)}$ are defined from equations (\ref{eq:Schur_complements_matrix}) and (\ref{eq:Forward_elimination_sources_matrix}). Obtaining the matrix $\vb*{\Delta}_k$ directly from equation (\ref{eq:Schur_complements_matrix}) requires to invert $\vb*{\Delta}_{k+1}$, perform two matrix multiplications and a subtraction of matrices. The inversion using LU factorization and each matrix multiplication require $O(N_{\text{fs}}^3)$ operations so it is desirable to reduce the number of matrix multiplications as much as possible. We can reduce the number of matrix multiplications in determining $\vb*{\Delta}_{k}$ to one if instead of computing $\vb*{\Delta}_{k+1}^{-1}$ we solve the matrix system of equations
\begin{align}
	\vb*{\Delta}_{k+1} \vb*{X}_{k+1} = \vb*{L}_{k+1},
	\label{eq:Forward_elimination_X}  
\end{align}
for $\vb*{X}_{k+1}$ and then obtain 
\begin{align}
	\vb*{\Delta}_k = \vb*{D}_k - \vb*{U}_{k}\vb*{X}_{k+1}, 
	\label{eq:Schur_complements_Fourier_collocation}
\end{align}
for $k=N_\xi-1, N_\xi-2, \ldots, 0$. Thus, obtaining $\vb*{\Delta}_k$ requires $O(N_{\text{fs}}^3)$ operations for solving equation (\ref{eq:Forward_elimination_X}) (using LU factorization) and also $O(N_{\text{fs}}^3)$ operations for applying (\ref{eq:Schur_complements_Fourier_collocation}). For computing the monoenergetic coefficients, the backward substitution step requires solving equation (\ref{eq:DKE_Forward_elimination}) for $k=0,1$ and $2$. Therefore, for $k\le 1$, it is convenient to store $\vb*{\Delta}_{k+1}$ in the factorized LU form obtained when equation (\ref{eq:Forward_elimination_X}) was solved for $\vb*{X}_{k+1}$. The matrix $\vb*{\Delta}_{0}$ will be factorized later, during the backward substitution step.

Similarly to what is done to obtain $\vb*{\Delta}_k$, to compute $\vb*{\sigma}^{(k)}$ we first solve 
\begin{align}
	\vb*{\Delta}_{k+1} \vb*{y} = \vb*{\sigma}^{(k+1)}
	\label{eq:Forward_elimination_matrix_y_system}
\end{align}
for $\vb*{y}$ and then compute
\begin{align}
	\vb*{\sigma}^{(k)} & = \vb*{s}^{(k)} - \vb*{U}_{k}  \vb*{y},
	\label{eq:Forward_elimination_sources_matrix_y}
\end{align}
for $k\ge 0$. Recall that none of the source terms $s_1$, $s_2$ and $s_3$ defined by (\ref{eq:DKE_Sources}) have Legendre modes greater than 2. Specifically, equation (\ref{eq:Forward_elimination_sources_matrix}) implies $\vb*{\sigma}_1^{(k)}, \vb*{\sigma}_3^{(k-1)} = 0$ for $k\ge 3$ and also $\vb*{\sigma}_1^{(2)} = \vb*{s}_1^{(2)}$, $\vb*{\sigma}_3^{(1)} = \vb*{s}_3^{(1)}$. Thus, we only have to solve equation (\ref{eq:Forward_elimination_matrix_y_system}) and apply (\ref{eq:Forward_elimination_sources_matrix_y}) to obtain $\{\vb*{\sigma}_1^{(k)}\}_{k=0}^{1}$ and $\vb*{\sigma}_3^{(0)}$. As $\{\vb*{\Delta}_{k+1}\}_{k=0}^{1}$ are already LU factorized, solving equation (\ref{eq:Forward_elimination_matrix_y_system}) and then applying (\ref{eq:Forward_elimination_sources_matrix_y}) requires $O(N_{\text{fs}}^2)$ operations and its contribution to the arithmetic complexity of the algorithm is subdominant with respect to the $O(N_{\text{fs}}^3)$ operations required to compute $\vb*{\Delta}_k$.

For the backward substitution, we first note that solving the matrix version of equation (\ref{eq:DKE_Forward_elimination}) to obtain $\vb*{f}^{(0)}$ requires $O(N_{\text{fs}}^3)$ operations, as $\vb*{\Delta}_0$ has not been LU factorized during the forward elimination. On the other hand, obtaining the remaining modes  $\{\vb*{f}^{(k)}\}_{k=1}^{2}$, requires $O(N_{\text{fs}}^2)$ operations. As the resolution of the matrix system of equations (\ref{eq:Forward_elimination_X}) and the matrix multiplication in (\ref{eq:Schur_complements_Fourier_collocation}) must be done $N_\xi$ times, solving equation (\ref{eq:DKE_Legendre_expansion_Fourier_collocation}) by this method requires $O(N_\xi N_{\text{fs}}^3)$ operations.

In what concerns to memory resources, as we are only interested in the Legendre modes $0$, $1$ and $2$, it is not necessary to store in memory all the matrices $\vb*{L}_k$, $\vb*{D}_k$, $\vb*{U}_k$ and $\vb*{\Delta}_k$. Instead, we store solely $\vb*{L}_k$, $\vb*{U}_k$ and $\vb*{\Delta}_k$ (in LU form) for $k=0,1,2$. For the intermediate steps we just need to use some auxiliary matrices $\vb*{L}$, $\vb*{D}$, $\vb*{U}$, $\vb*{\Delta}$ and $\vb*{X}$ of size $N_{\text{fs}}$. This makes the amount of memory required by {\MONKES} independent of $N_\xi$, being of order $N_{\text{fs}}^2$.
\begin{algorithm} 
	\caption{Block tridiagonal solution algorithm implemented in {\MONKES}.}\label{alg:MONKES_BTD}		
	\textbf{1. Forward elimination:}
	\begin{algorithmic}
		\State{
			$\vb*{L}\gets \vb*{L}_{N_\xi}$} \Comment{Starting value for $\vb*{L}$}
		\State{$\vb*{\Delta}\gets \vb*{D}_{N_\xi}$} \Comment{Starting value for $\vb*{\Delta}$} 
		\State{Solve $\vb*{\Delta} \vb*{X}=\vb*{L}$} \Comment{Compute $\vb*{X}_{N_\xi}$ stored in $\vb*{X}$}
		
		\For{$k=N_\xi-1$ \textbf{to} $0$}
		\State{$\vb*{L}\gets \vb*{L}_{k}$} 
		\Comment{Construct $\vb*{L}_k$ stored in $\vb*{L}$}
		\State{$\vb*{D}\gets \vb*{D}_{k}$} 
		\Comment{Construct $\vb*{D}_k$ stored in $\vb*{D}$}
		\State{$\vb*{U}\gets \vb*{U}_{k}$} 
		\Comment{Construct $\vb*{U}_k$ stored in $\vb*{U}$}
		\State{$\vb*{\Delta}\gets \vb*{D} - \vb*{U} \vb*{X}$} \Comment{Construct $\vb*{\Delta}_k$ stored in $\vb*{\Delta}$}
		\State{\textbf{if} $k>0$:\, Solve $\vb*{\Delta} \vb*{X}=\vb*{L}$} \Comment{Compute $\vb*{X}_{k}$ stored \qquad \qquad\qquad\qquad\qquad in $\vb*{X}$ for next iteration}
		
		\If{$k\le2$}\Comment{Save required matrices }
		\State{\textbf{if} $k=0$: $\vb*{L}_{k}\gets \vb*{L}$  	  \Comment{Save $\{\vb*{L}_k\}_{k=1}^{2}$}} 
		\State{$\vb*{U}_{k}\gets \vb*{U}$ 	\Comment{Save $\{\vb*{U}_k\}_{k=0}^{2}$  } }
		\State{$\vb*{\Delta}_{k}\gets \vb*{\Delta}$}    \Comment{Save $\{\vb*{\Delta}_k\}_{k=0}^{2}$}
		\EndIf
		\EndFor
		
		\State{}
		
		\For{$k=1$ to $0$}
		\State{Solve $\vb*{\Delta}_{k+1}\vb*{y}_1 = \vb*{\sigma}_1^{(k+1)} $} 
		
		\State{\textbf{if} $k=0$: Solve $\vb*{\Delta}_{k+1}\vb*{y}_3 = \vb*{\sigma}_3^{(k+1)} $}
		
		\State{$\vb*{\sigma}_1^{(k)} \gets \vb*{s}_1^{(k)} - \vb*{U}_k \vb*{y}_1$} \Comment{Construct $\vb*{\sigma}_1^{(k)}$ }
		\State{\textbf{if} $k=0$:  $\vb*{\sigma}_3^{(0)} \gets - \vb*{U}_0 \vb*{y}_3$} \Comment{Construct $\vb*{\sigma}_3^{(0)}$ }
		
		\EndFor

		\State{}
	\end{algorithmic}
	
	\textbf{2. Backward substitution:}
	\begin{algorithmic}
		\State{Solve $\vb*{\Delta}_0 \vb*{f}^{(0)} = \vb*{\sigma}^{(0)}$}
		\For{$k=1$ \textbf{to} $2$}    	
		\State{
			Solve $\vb*{\Delta}_k\vb*{f}^{(k)} = 
			\vb*{\sigma}^{(k)} - \vb*{L}_{k} \vb*{f}^{(k-1)} $
		}
		\EndFor
	\end{algorithmic}
\end{algorithm}

To summarize, the pseudocode of the implementation of the algorithm in {\MONKES} is given in Algorithm \ref{alg:MONKES_BTD}. In the first loop from $k=N_\xi-1$ to $k=0$ we construct and save only the matrices $\{\vb*{L}_k,\vb*{U}_k, \vb*{\Delta}_k\}_{k=0}^{2}$. At this point the matrices $\{\vb*{\Delta}_k\}_{k=1}^{2}$ are factorized in LU form. In the second loop, the sources $\{\vb*{\sigma}_1^{(k)}\}_{k=0}^{1}$ and $\vb*{\sigma}_3^{(0)}$ are computed and saved for the backward substitution. Finally, the backward substitution step is applied. For solving $\vb*{\Delta}_0 \vb*{f}^{(0)} = \vb*{\sigma}^{(0)}$ we have to perform the LU factorization of $\vb*{\Delta}_0$ (just for one of the two source terms) and then solve for $\vb*{f}^{(0)}$. For the remaining modes, the LU factorizations of $\{\vb*{\Delta}_k\}_{k=1}^{2}$ are reused to solve for $\{\vb*{f}^{(k)}\}_{k=1}^{2}$.

Once we have solved equation (\ref{eq:DKE_Legendre_expansion_Fourier_collocation}) for $\vb*{f}^{(0)}$, $\vb*{f}^{(1)}$ and $\vb*{f}^{(2)}$, the integrals of the flux-surface average operation involved in the monoenergetic coefficients (\ref{eq:Gamma_11_Legendre}), (\ref{eq:Gamma_31_Legendre}), (\ref{eq:Gamma_13_Legendre}) and (\ref{eq:Gamma_33_Legendre}), are conveniently computed using the trapezoidal rule, which for periodic analytic functions has geometric convergence \cite{Trapezoidal}. In section \ref{sec:Results_Benchmark} we will see that despite the cubic scaling in $N_{\text{fs}}$ of the arithmetical complexity of the algorithm, it is possible to obtain fast and accurate calculations of the monoenergetic geometric coefficients at low collisionality (and in particular $\widehat{D}_{31}$) in a single core. The reason behind this is that in the asymptotic relation $O(N_{\text{fs}}^3)\sim C_{\text{alg}} N_{\text{fs}}^3$, the constant $C_{\text{alg}}$ is small enough to allow $N_{\text{fs}}$ to take a sufficiently high value to capture accurately the spatial dependence of the distribution function without increasing much the wall-clock time. 

The algorithm is implemented in the new code {\MONKES}, written in Fortran language. The matrix inversions and multiplications are computed using the linear algebra library \texttt{LAPACK} \cite{lapack99}.
\section{Code performance and benchmark}
\label{sec:Results_Benchmark}
In this section we will demonstrate that {\MONKES} provides fast and accurate calculations of the monoenergetic coefficients from low ($\hat{\nu}=10^{-5}$ $\text{m}^{-1}$) to high collisionality\footnote{In this context ``accurate at high collisionality'' means that the drift-kinetic equation (\ref{eq:DKE}) is solved accurately.} ($\hat{\nu}=3\cdot10^{2}$ $\text{m}^{-1}$).

In subsection \ref{subsec:Convergence} we will see that for a correct calculation of the monoenergetic coefficients for $\hat{\nu}\ge 10^{-5}$ m, $N_{\text{fs}} \lesssim  2000$ and $N_\xi \lesssim 200$ are required. In subsection \ref{subsec:Performance} it is shown that for these resolutions {\MONKES} produces fast calculations in a single processor. Finally, in subsection \ref{subsec:Benchmark} the coefficients computed with {\MONKES} will be benchmarked with {\DKES} and {\texttt{SFINCS}}. As a result of the benchmarking, we will conclude that {\MONKES} calculations are accurate.

\subsection{Convergence of monoenergetic coefficients at low collisionality}
\label{subsec:Convergence}
In low collisionality regimes, convection is dominant with respect to diffusion. As equation (\ref{eq:DKE}) is singularly perturbed with respect to $\hat{\nu}$, its solution possesses internal boundary layers in $\xi$. These boundary layers appear at the interfaces between different classes of trapped particles. At these regions of phase space, collisions are no longer subdominant with respect to advection. Besides, at these regions, the poloidal $\vb*{E}\times\vb*{B}$ precession from equation (\ref{eq:DKE}) can produce the chaotic transition of collisionless particles from one class to another due to separatrix crossing mechanisms \cite{Cary_Separatrix_Crossing, dherbemont2022}. The existence of these localized regions with large $\xi$ gradients demands a high number of Legendre modes $N_\xi$, explaining the difficulty to obtain fast and accurate solutions to equation (\ref{eq:DKE}) at low collisionality. 

In this subsection we will select resolutions $N_\theta$, $N_\zeta$ and $N_\xi$ for which {\MONKES} provides accurate calculations of the monoenergetic coefficients in a wide range of collisionalities. For this, we will study how the monoenergetic coefficients computed by {\MONKES} converge with $N_\theta$, $N_\zeta$ and $N_\xi$ at low collisionality. From the point of view of numerical analysis, the need for large values of $N_\xi$ is due to the lack of diffusion along $\xi$ in equation (\ref{eq:DKE}). Hence, if {\MONKES} is capable of producing fast and accurate calculations at low collisionality, it will also produce fast and accurate calculations at higher collisionalities. 

For the convergence study, we select three different magnetic configurations at a single flux surface. Two of them correspond to configurations of W7-X: EIM and KJM. The third one corresponds to the new QI ``flat mirror'' \cite{velasco2023robust} configuration CIEMAT-QI \cite{Sanchez_2023}. The calculations are done for the $1/\nu$ (cases with $\widehat{E}_r=0$) and $\sqrt{\nu}$-$\nu$ regimes \cite{dherbemont2022} (cases with $\widehat{E}_r\ne 0$) at the low collisionality value $\hat{\nu}=10^{-5}$ m. In table \ref{tab:Convergence_cases} the cases considered are listed, including their correspondent values of $\widehat{E}_r:= \widehat{E}_\psi\dv*{\psi}{r}$. We have denoted $r = a \sqrt{\psi/\psi_{\text{lcfs}}}$ and, in this context, $a$ is the minor radius of the device\footnote{{\DKES} uses $r$ as radial coordinate instead of $\psi$. The quantities $\hat{\nu}$ and $\widehat{E}_r$ are denoted respectively \texttt{CMUL} and \texttt{EFIELD} in the code {\DKES}.}.
\begin{table}[h]
	\centering
	\begin{tabular}{@{}lccc@{}}
		\toprule
		Configuration & $\psi/\psi_{\text{lcfs}}$ & $\hat{\nu}$ $[\text{m}^{-1}]$ & $\widehat{E}_r$  $[\text{kV}\cdot\text{s}/\text{m}^2]$   \\ \midrule
		W7X-EIM       & 0.200                     & $10^{-5}$   & 0 \\
		W7X-EIM       & 0.200                     & $10^{-5}$   & $3\cdot10^{-4}$ \\
		W7X-KJM       & 0.204                     & $10^{-5}$   & 0 \\
		W7X-KJM       & 0.204                     & $10^{-5}$   & $3\cdot10^{-4}$ \\ 
		CIEMAT-QI     & 0.250                     & $10^{-5}$   & 0       \\
		CIEMAT-QI     & 0.250                     & $10^{-5}$   & $10^{-3}$       \\
		\bottomrule
	\end{tabular}
	\caption{Cases considered in the convergence study of monoenergetic coefficients and values of $(\hat{\nu},\widehat{E}_r)$.}
	\label{tab:Convergence_cases}
\end{table}

In order to select the triplets $(N_\theta,N_\zeta, N_\xi)$ for sufficiently accurate calculations of $\widehat{D}_{31}$, we need to specify when we will consider that a computation has converged. For each case of table \ref{tab:Convergence_cases} we will proceed in the same manner. First, we plot the coefficients $\widehat{D}_{ij}$ as functions of the number of Legendre modes in a sufficiently wide interval. For each value of $N_\xi$, the selected spatial resolutions $N_\theta$ and $N_\zeta$ are large enough so that increasing them varies the monoenergetic coefficients in less than a 1\%. We will say that these calculations are ``spatially converged''. Since, typically, the most difficult coefficient to calculate is the bootstrap current coefficient, we will select the resolutions so that $\widehat{D}_{31}$ is accurately computed. From the curve of (spatially converged) $\widehat{D}_{31}$ as a function of $N_\xi$ we define our converged reference value, which we denote by $\widehat{D}_{31}^{\text{r}}$, as the converged calculation to three significant digits. From this converged reference value we will define two regions. A first region
\begin{align}
	\mathcal{R}_{\epsilon}:=
	\left[
	(1-\epsilon/100)\widehat{D}_{31}^{\text{r}}, (1+\epsilon/100)\widehat{D}_{31}^{\text{r}} 
	\right]
\end{align} 
for calculations that deviate less than or equal to an $\epsilon$\% with respect to $\widehat{D}_{31}^{\text{r}}$. This interval will be used for selecting the resolutions through the following convergence criteria. We say that, for fixed $(N_\theta,N_\zeta,N_\xi)$ and $\epsilon$, a calculation $\widehat{D}_{31}\in\mathcal{R}_{\epsilon}$ is sufficiently converged if two conditions are satisfied 
\begin{enumerate}
	\item Spatially converged calculations with $N_\xi'\ge N_\xi$ belong to $\mathcal{R}_{\epsilon}$.
	\item Increasing $N_\theta$ and $N_\zeta$ while keeping $N_\xi$ constant produces calculations which belong to $\mathcal{R}_{\epsilon}$.
\end{enumerate}
Condition (i) is used to select the number of Legendre modes $N_\xi$ and condition (ii) is used to select the values of $N_\theta$ and $N_\zeta$ once $N_\xi$ is fixed. 

Additionally, we define a second interval 
\begin{align}
	\mathcal{A}_{\epsilon}:=
	\left[
	\widehat{D}_{31}^{\text{r}}-\epsilon, \widehat{D}_{31}^{\text{r}}+\epsilon 
	\right]
\end{align}
to distinguish which calculations are at a distance smaller than or equal to $\epsilon$ from $\widehat{D}_{31}^{\text{r}}$. The reason to have two different regions is that for stellarators close to QI, the relative convergence criteria can become too demanding (the smaller $\widehat{D}_{31}^{\text{r}}$ is, the narrower $\mathcal{R}_{\epsilon}$ becomes). Nevertheless, for optimizing QI configurations, it is sufficient to ensure that $|\widehat{D}_{31}|$ is sufficiently small. If the absolute error is much smaller than a value of $|\widehat{D}_{31}|$ that can be considered sufficiently small, the calculation is converged for optimization purposes. We will use this interval for two purposes: first to give a visual idea of how narrow $\mathcal{R}_{\epsilon}$ becomes. Second, to show that if $\mathcal{R}_{\epsilon}$ is very small, it is easier to satisfy an absolute criteria than a relative one. 

\begin{figure*}[h]
	\centering
	\begin{subfigure}[t]{0.32\textwidth}
		\tikzsetnextfilename{Convergence-Legendre-W7X-EIM-s0200-Er-0-D11}
		\includegraphics{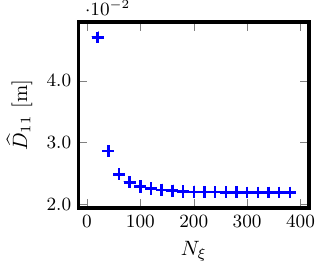}
		\caption{}
		\label{subfig:D11_Convergence_W7X_EIM_Er_0_Legendre}
	\end{subfigure}
	\begin{subfigure}[t]{0.32\textwidth}
		\tikzsetnextfilename{Convergence-Legendre-W7X-EIM-s0200-Er-0-D33}
		\includegraphics{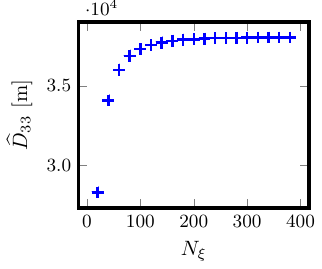}
		\caption{}
		\label{subfig:D33_Convergence_W7X_EIM_Er_0_Legendre}
	\end{subfigure}
	
	\begin{subfigure}[t]{0.32\textwidth}
		\tikzsetnextfilename{Convergence-Legendre-W7X-EIM-s0200-Er-0-D31-Detail}
		\includegraphics{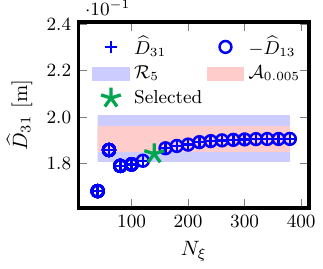}
		\caption{}
		\label{subfig:D31_convergence_Legendre_W7X_EIM_0200_Erho_0_Detail}
	\end{subfigure}
	\begin{subfigure}[t]{0.32\textwidth}
		\tikzsetnextfilename{Convergence-theta-zeta-W7X-EIM-s0200-Er-0-D31}
		\includegraphics{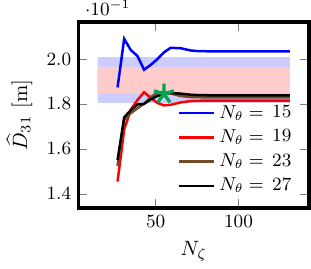}
		\caption{}
		\label{subfig:D31_convergence_theta_zeta_W7X_EIM_0200_Erho_0}
	\end{subfigure}
	\caption{Convergence of monoenergetic coefficients with the number of Legendre modes $N_\xi$ for W7X-EIM at the surface labelled by $\psi/\psi_{\text{lcfs}}=0.200$, for $\hat{\nu}(v)=10^{-5}$ $\text{m}^{-1}$ and $\widehat{E}_r(v)=0$ $\text{kV}\cdot\text{s}/\text{m}^2$.}
	\label{fig:Convergence_W7X_EIM_Er_0}
\end{figure*}

Figure \ref{fig:Convergence_W7X_EIM_Er_0} shows the convergence of monoenergetic coefficients with the number of Legendre modes for W7-X EIM when $\widehat{E}_r=0$. From figures \ref{subfig:D11_Convergence_W7X_EIM_Er_0_Legendre} and \ref{subfig:D33_Convergence_W7X_EIM_Er_0_Legendre} we see that the radial transport and parallel conductivity coefficients converge monotonically with $N_\xi$. On the other hand, the bootstrap current coefficient is more difficult to converge as can be seen on figure \ref{subfig:D31_convergence_Legendre_W7X_EIM_0200_Erho_0_Detail}. As a sanity check, the fulfilment of the Onsager symmetry relation $\widehat{D}_{31}= - \widehat{D}_{13}$ is included. The converged reference value $\widehat{D}_{31}^{\text{r}}$ is the spatially converged calculation for $N_\xi=380$. Defining a region of relative convergence of $\epsilon=5\%$, allows to select a resolution of $N_\xi=140$ Legendre modes to satisfy condition (i). The selection is indicated with a five-pointed green star. Note that for this case, an absolute deviation of $0.005$ m from $\widehat{D}_{31}^{\text{r}}$ is slightly more demanding than the relative condition. This absolute deviation is selected as the 5\% of $\widehat{D}_{31}\sim 0.1$ m, which can be considered a small value of $\widehat{D}_{31}$. From figure \ref{subfig:D31_convergence_theta_zeta_W7X_EIM_0200_Erho_0} we choose the resolutions $(N_\theta,N_\zeta)=(23,55)$ to satisfy convergence condition (ii). 

\begin{figure*}[]
	\centering
	\begin{subfigure}[t]{0.32\textwidth}
		\tikzsetnextfilename{Convergence-Legendre-W7X-EIM-s0200-Er-3e-4-D11}
		\includegraphics{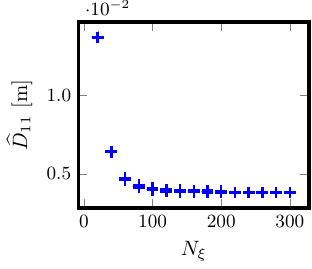}
		\caption{}
		\label{subfig:D11_convergence_Legendre_W7X_EIM_0200_Erho_3e-4}
	\end{subfigure}
	\begin{subfigure}[t]{0.32\textwidth}
		\tikzsetnextfilename{Convergence-Legendre-W7X-EIM-s0200-Er-3e-4-D33}
		\includegraphics{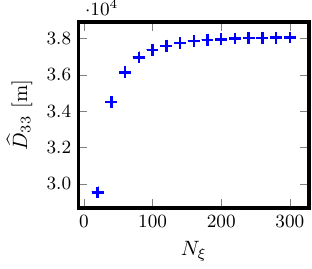}
		\caption{}
		\label{subfig:D33_convergence_Legendre_W7X_EIM_0200_Erho_3e-4}
	\end{subfigure}

	\begin{subfigure}[t]{0.32\textwidth}
		\tikzsetnextfilename{Convergence-Legendre-W7X-EIM-s0200-Er-3e-4-D31-Detail}
		\includegraphics{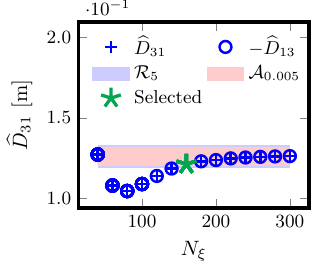}
		\caption{}
		\label{subfig:D31_convergence_Legendre_W7X_EIM_0200_Erho_3e-4_Detail}
	\end{subfigure}
	\begin{subfigure}[t]{0.32\textwidth}
		\tikzsetnextfilename{Convergence-theta-zeta-W7X-EIM-s0200-Er-3e-4-D31}
		\includegraphics{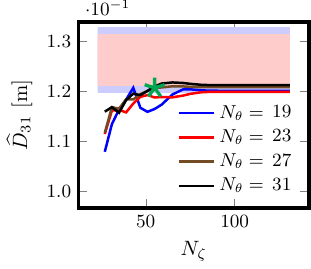}
		\caption{}
		\label{subfig:D31_convergence_theta_zeta_W7X_EIM_0200_Erho_3e-4_Detail}
	\end{subfigure}

	\caption{Convergence of monoenergetic coefficients with the number of Legendre modes $N_\xi$ for W7X-EIM at the surface labelled by $\psi/\psi_{\text{lcfs}}=0.200$, for $\hat{\nu}(v)=10^{-5}$ $\text{m}^{-1}$ and $\widehat{E}_r=3\cdot 10^{-4}$ $\text{kV}\cdot\text{s}/\text{m}^2$.}
	\label{fig:Convergence_W7X_EIM_Er_3e-4}
\end{figure*}

The case of W7-X EIM with $\widehat{E}_r\ne 0$ is shown in figure \ref{fig:Convergence_W7X_EIM_Er_3e-4}. We note from figure \ref{subfig:D31_convergence_Legendre_W7X_EIM_0200_Erho_3e-4_Detail} that obtaining sufficiently converged results for the region $\mathcal{R}_{5}$ is more difficult than in the case without radial electric field. For this case, the sizes of the intervals $\mathcal{A}_{0.005}$ and $\mathcal{R}_{5}$ are almost the same. This is in part due to the fact that the $\widehat{D}_{31}$ coefficient is smaller in absolute value and thus, the region $\mathcal{R}_{5}$ is narrower. We select $N_\xi=160$ to satisfy condition (i). The selection $(N_\theta,N_\zeta)=(27,55)$ satisfies condition (ii) as shown in figure \ref{subfig:D31_convergence_theta_zeta_W7X_EIM_0200_Erho_3e-4_Detail}. 

\begin{figure*}[t]
	\centering
	\begin{subfigure}[t]{0.32\textwidth}
		\tikzsetnextfilename{Convergence-Legendre-W7X-KJM-s0204-Er-0-D11}
		\includegraphics{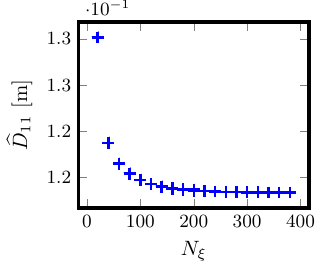}
		\caption{}
		\label{subfig:D11_convergence_Legendre_W7X_KJM_0204_Erho_0}
	\end{subfigure}
	\begin{subfigure}[t]{0.32\textwidth}
		\tikzsetnextfilename{Convergence-Legendre-W7X-KJM-s0204-Er-0-D33}
		\includegraphics{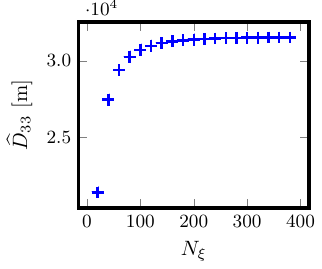}
		\caption{}
		\label{subfig:D33_convergence_Legendre_W7X_KJM_0204_Erho_0}
	\end{subfigure}

	\begin{subfigure}[t]{0.32\textwidth}
		\tikzsetnextfilename{Convergence-Legendre-W7X-KJM-s0204-Er-0-D31-Detail}
		\includegraphics{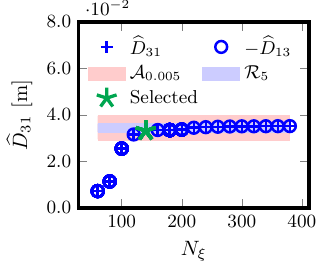}
		\caption{}
		\label{subfig:D31_convergence_Legendre_W7X_KJM_0204_Erho_0_Detail}
	\end{subfigure}
	\begin{subfigure}[t]{0.32\textwidth}
		\tikzsetnextfilename{Convergence-theta-zeta-W7X-KJM-s0204-Er-0-D31}
		\includegraphics{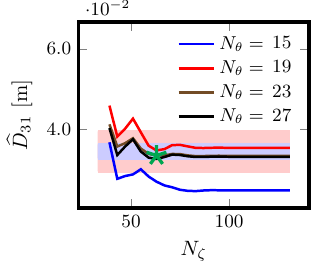}
		\caption{}
		\label{subfig:D31_convergence_theta_zeta_W7X_KJM_0204_Erho_0}
	\end{subfigure}
	\caption{Convergence of monoenergetic coefficients with the number of Legendre modes $N_\xi$ for W7X-KJM at the surface labelled by $\psi/\psi_{\text{lcfs}}=0.204$, for $\hat{\nu}(v)=10^{-5}$ $\text{m}^{-1}$ and $\hat{E}_r(v)=0$ $\text{kV}\cdot\text{s}/\text{m}^2$.}
	\label{fig:Convergence_W7X_KJM_Er_0}
\end{figure*}

The convergence curves for the case of W7-X KJM when $\widehat{E}_r=0$ are shown in figure \ref{fig:Convergence_W7X_KJM_Er_0}. Due to the smallness of $\widehat{D}_{31}^{\text{r}}$, the amplitude of the region $\mathcal{R}_{5}$ is much narrower than in the EIM case, being of order $10^{-3}$. It is so narrow that the absolute value region $\mathcal{A}_{0.005}$ contains the relative convergence region. It is shown in figure \ref{subfig:D31_convergence_Legendre_W7X_KJM_0204_Erho_0_Detail} that taking $N_\xi=140$ is sufficient to satisfy condition (i). According to the convergence curves plotted in figure \ref{subfig:D31_convergence_theta_zeta_W7X_KJM_0204_Erho_0}, selecting $(N_\theta,N_\zeta)=(23,63)$ ensures satisfying condition (ii). 

\begin{figure*}[t]
	\centering
	\begin{subfigure}[t]{0.32\textwidth}
		\tikzsetnextfilename{Convergence-Legendre-W7X-KJM-s0204-Er-3e-4-D11}
		\includegraphics{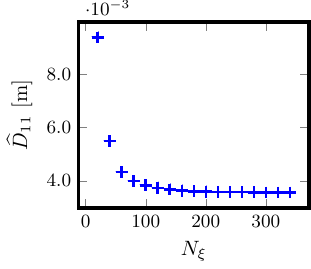}
		\caption{}
		\label{subfig:D11_convergence_Legendre_W7X_KJM_0204_Erho_3e-4}
	\end{subfigure}
	\begin{subfigure}[t]{0.32\textwidth}
		\tikzsetnextfilename{Convergence-Legendre-W7X-KJM-s0204-Er-3e-4-D33}
		\includegraphics{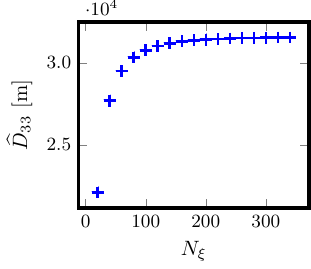}
		\caption{}
		\label{subfig:D33_convergence_Legendre_W7X_KJM_0204_Erho_3e-4}
	\end{subfigure}

	\begin{subfigure}[t]{0.32\textwidth}
		\tikzsetnextfilename{Convergence-Legendre-W7X-KJM-s0204-Er-3e-4-D31-Detail}
		\includegraphics{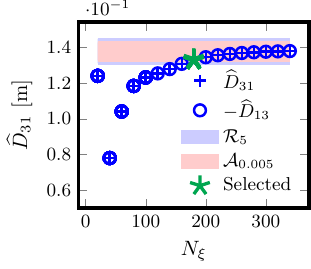}
		\caption{}
		\label{subfig:D31_convergence_Legendre_W7X_KJM_0204_Erho_3e-4_Detail}
	\end{subfigure}
	\begin{subfigure}[t]{0.32\textwidth}
		\tikzsetnextfilename{Convergence-theta-zeta-W7X-KJM-s0204-Er-3e-4-D31}
		\includegraphics{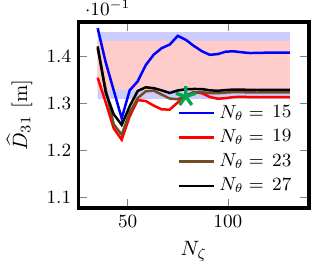}
		\caption{}
		\label{subfig:D31_convergence_theta_zeta_W7X_KJM_0204_Erho_3e-4_Detail}
	\end{subfigure}

	\caption{Convergence of monoenergetic coefficients with the number of Legendre modes $N_\xi$ for W7X-KJM at the surface labelled by $\psi/\psi_{\text{lcfs}}=0.204$, for $\hat{\nu}(v)=10^{-5}$ $\text{m}^{-1}$ and $\widehat{E}_r(v)=3\cdot 10^{-4}$ $\text{kV}\cdot\text{s}/\text{m}^2$.}
	\label{fig:Convergence_W7X_KJM_Er_3e-4}
\end{figure*}

The case of W7-X KJM for finite $\widehat{E}_r$ is shown in figure \ref{fig:Convergence_W7X_KJM_Er_3e-4}. The selection of $N_\xi=180$ Legendre modes, indicated in figure \ref{subfig:D31_convergence_Legendre_W7X_KJM_0204_Erho_3e-4_Detail}, satisfies convergence condition (i). As shown in figure \ref{subfig:D31_convergence_theta_zeta_W7X_KJM_0204_Erho_3e-4_Detail}, condition (ii) is satisfied by the selection $(N_\theta, N_\zeta)=(19,79)$.



\begin{figure*}[t]
	\centering
	\begin{subfigure}[t]{0.32\textwidth}
		\tikzsetnextfilename{Convergence-Legendre-CIEMAT-QI-s0250-Er-0-D11}
		\includegraphics{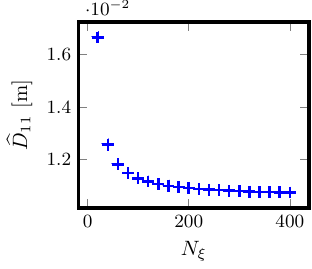}
		\caption{}
		\label{subfig:D11_convergence_Legendre_CIEMAT_QI_0250_Erho_0}
	\end{subfigure}
	\begin{subfigure}[t]{0.32\textwidth}
		\tikzsetnextfilename{Convergence-Legendre-CIEMAT-QI-s0250-Er-0-D33}
		\includegraphics{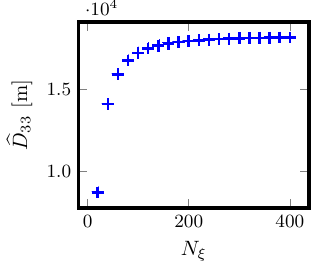}
		\caption{}
		\label{subfig:D33_convergence_Legendre_CIEMAT_QI_0250_Erho_0}
	\end{subfigure}

	\begin{subfigure}[t]{0.32\textwidth}
		\tikzsetnextfilename{Convergence-Legendre-CIEMAT-QI-s0250-Er-0-D31-Detail}
		\includegraphics{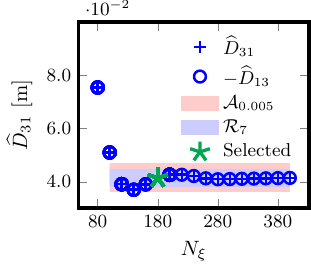}
		\caption{}
		\label{subfig:D31_convergence_Legendre_CIEMAT_QI_0250_Erho_0_Detail}
	\end{subfigure}
	\begin{subfigure}[t]{0.32\textwidth}
		\tikzsetnextfilename{Convergence-theta-zeta-CIEMAT-QI-s0250-Er-0-D31}
		\includegraphics{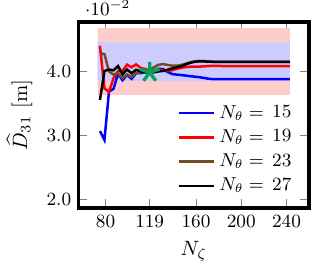}
        \caption{}
        \label{subfig:D31_convergence_theta_zeta_CIEMAT_QI_0250_Erho_0}
	\end{subfigure}

	\caption{Convergence of monoenergetic coefficients with the number of Legendre modes $N_\xi$ for CIEMAT-QI at the surface labelled by $\psi/\psi_{\text{lcfs}}=0.25$, for $\hat{\nu}(v)=10^{-5}$ $\text{m}^{-1}$ and $\widehat{E}_r(v)=0$ $\text{kV}\cdot\text{s}/\text{m}^2$.}
	\label{fig:Convergence_CIEMAT_QI_Er_0}
\end{figure*}
\begin{figure*}[t]
	\centering
	\begin{subfigure}[t]{0.32\textwidth}
		\tikzsetnextfilename{Convergence-Legendre-CIEMAT-QI-s0250-Er-1e-3-D11}
		\includegraphics{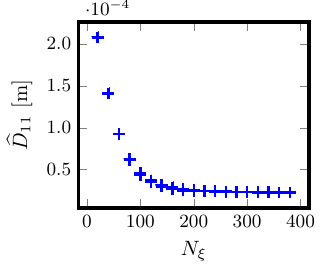}
		\caption{}
		\label{subfig:D11_convergence_Legendre_CIEMAT_QI_0250_Erho_1e-3}
	\end{subfigure}
	\begin{subfigure}[t]{0.32\textwidth}
		\tikzsetnextfilename{Convergence-Legendre-CIEMAT-QI-s0250-Er-1e-3-D33}
		\includegraphics{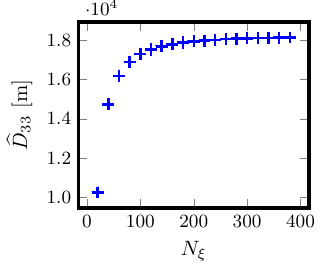}
		\caption{}
		\label{subfig:D33_convergence_Legendre_CIEMAT_QI_0250_Erho_1e-3}
	\end{subfigure}
	
	\begin{subfigure}[t]{0.32\textwidth}
		\tikzsetnextfilename{Convergence-Legendre-CIEMAT-QI-s0250-Er-1e-3-D31-Detail}
		\includegraphics{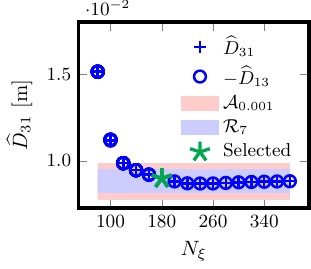}
		\caption{}
		\label{subfig:D31_convergence_Legendre_CIEMAT_QI_0250_Erho_1e-3_Detail}
	\end{subfigure}
	\begin{subfigure}[t]{0.32\textwidth}
		\tikzsetnextfilename{Convergence-theta-zeta-CIEMAT-QI-s0250-Er-1e-3-D31}
		\includegraphics{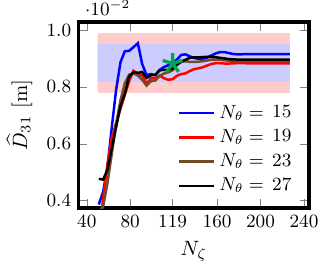}
		\caption{}
		\label{subfig:D31_convergence_theta_zeta_CIEMAT_QI_0250_Erho_1e-3}
	\end{subfigure}
	\caption{Convergence of monoenergetic coefficients with the number of Legendre modes $N_\xi$ for CIEMAT-QI at the surface labelled by $\psi/\psi_{\text{lcfs}}=0.25$, for $\hat{\nu}(v)=10^{-5}$ $\text{m}^{-1}$ and $\widehat{E}_r(v)=10^{-3}$ $\text{kV}\cdot\text{s}/\text{m}^2$.}
	\label{fig:Convergence_CIEMAT_QI_Er_1e-3}
\end{figure*}

The convergence of monoenergetic coefficients for CIEMAT-QI without $\widehat{E}_r$ is shown in figure \ref{fig:Convergence_CIEMAT_QI_Er_0}. Note that as in the W7-X KJM case at this regime, the region of absolute error $\mathcal{A}_{0.005}$ is bigger than the relative one. As the monoenergetic coefficients are smaller, we relax the relative convergence parameter to $\epsilon=7\%$. In figure \ref{subfig:D31_convergence_Legendre_CIEMAT_QI_0250_Erho_0_Detail} we see that the region of 7\% of deviation $\mathcal{R}_{7}$ is quite narrow and that selecting $N_\xi=180$ satisfies condition (i). To satisfy condition (ii), we choose the resolutions $(N_\theta,N_\zeta)=(15,119)$ as shown in figure \ref{subfig:D31_convergence_theta_zeta_CIEMAT_QI_0250_Erho_0}.

Finally, the case of CIEMAT-QI with $\widehat{E}_r\ne 0$ is shown in figure \ref{fig:Convergence_CIEMAT_QI_Er_1e-3}. Looking at figure \ref{subfig:D31_convergence_Legendre_CIEMAT_QI_0250_Erho_1e-3_Detail} we can check that taking $N_\xi=180$ satisfies condition (i) for the region $\mathcal{R}_7$ of 7\% of deviation. In this case, the region of absolute error $\mathcal{A}_{0.001}$ is five times smaller than in the rest of cases and is still bigger than the relative error region. As shown in figure \ref{subfig:D31_convergence_theta_zeta_CIEMAT_QI_0250_Erho_1e-3}, the selection $(N_\theta,N_\zeta)=(15,119)$ satisfies condition (ii).

\subsection{Code performance}\label{subsec:Performance}
 In this subsection we will compare {\MONKES} and {\DKES} performance in terms of the wall-clock time and describe {\MONKES} scaling properties. For the wall-clock time comparison, a convergence study (similar to the one explained in subsection \ref{subsec:Convergence}) has been carried out with {\DKES} on \ref{sec:Appendix_DKES_Bounds}. This convergence study is done to compare the wall-clock times between {\MONKES} and {\DKES} for the same level of relative convergence with respect to $\widehat{D}_{31}^{\text{r}}$. The comparison is displayed in table \ref{tab:DKES_MONKES_Comparison} along with the minimum number of Legendre modes for which {\DKES} results satisfy convergence condition (i). In all six cases, {\MONKES} is faster than {\DKES} despite using more Legendre modes. Even for W7-X EIM, in which we have taken $N_\xi =40$ for {\DKES} calculations with finite $\widehat{E}_r$, {\MONKES} is $\sim 4$ times faster using almost four times the number of Legendre modes. For the W7-X EIM case without radial electric field, the speed-up is also of $ 4$. For the high mirror configuration, {\MONKES} is $\sim 20$ times faster than {\DKES} without $\widehat{E}_r$ and $\sim 10$ times faster than {\DKES} when $\widehat{E}_r \ne 0$. In the case of CIEMAT-QI, {\MONKES} is more than $\sim 13$ times faster than {\DKES} without radial electric field. In the case with finite $\widehat{E}_r$, {\MONKES} calculations are around 64 times faster than {\DKES} ones. One calculation of {\MONKES} takes less than a minute and a half and the same calculation with {\DKES} requires waiting for almost an hour and a half. We point out that the wall-clock times for all the calculations shown are those from one of the partitions of CIEMAT's cluster XULA. Specifically, partition number 2 has been used, whose nodes run with Intel Xeon Gold 6254 cores at 3.10 GHz. 

\begin{table}[h]
	\centering
	\begin{tabular}{lccc}
		\toprule
		Case   & $N_\xi^{\DKES}$ & $t_{\text{clock}}^{\DKES}$  [s] & $t_{\text{clock}}^{\MONKES}$  {[}s{]} \\ \midrule
		W7X-EIM $\widehat{E}_r=0$                                   & 80          &  90      &         22      \\
		W7X-EIM $\widehat{E}_r\ne 0$                                & {40}          &  {172}     &         35      \\ 
		W7X-KJM $\widehat{E}_r=0$                                   & 160         &  698     &         31      \\
		W7X-KJM $\widehat{E}_r\ne 0$                                & 60          &  421     &         47      \\
		CIEMAT-QI $\widehat{E}_r=0$                                 & 160         &  1060    &         76      \\
		CIEMAT-QI $\widehat{E}_r\ne 0$                              & 160         &  4990    &         76                    \\\bottomrule
	\end{tabular}
	\caption{Comparison between the wall-clock time of {\DKES} and {\MONKES}.}
	\label{tab:DKES_MONKES_Comparison}
\end{table}

\begin{figure}[h]
	\centering
	\begin{subfigure}[t]{0.35\textwidth}
		\tikzsetnextfilename{MONKES-Scaling-Legendre}
		\includegraphics{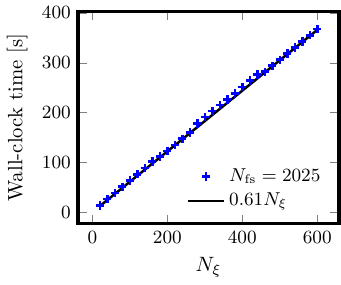} 
		\caption{}
		\label{subfig:MONKES_Scaling_Legendre}
	\end{subfigure}    
	\begin{subfigure}[t]{0.35\textwidth}
		\tikzsetnextfilename{MONKES-Scaling-Flux-Surface}
		\includegraphics{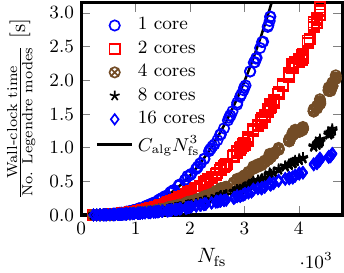}
		\caption{}
		\label{subfig:MONKES_Scaling_Nfs}
	\end{subfigure}
	\caption{Scaling of {\MONKES} wall-clock time. (a) Linear scaling with the number of Legendre modes for $N_{\text{fs}}=27\times 75 = 2025$ discretization points. (b) Cubic scaling with $N_{\text{fs}}$ for different number of cores used.}
	\label{fig:MONKES_Scaling}
\end{figure}

We next check that the arithmetic complexity of the algorithm described in section \ref{sec:Algorithm} holds in practice. The scaling of {\MONKES} with the number of Legendre modes $N_\xi$ and the number of points in which the flux-surface is discretized is shown in figure \ref{fig:MONKES_Scaling}. To demonstrate the linear scaling, the wall-clock time as a function of $N_\xi$ for $N_{\text{fs}}=2025$ points is represented in figure \ref{subfig:MONKES_Scaling_Legendre} and compared with the line of slope 0.61 seconds per Legendre mode. As can be seen in figure \ref{subfig:MONKES_Scaling_Nfs}, the wall-clock time (per Legendre mode) scales cubicly with the number of points in which the flux-surface is discretized $N_{\text{fs}}$. As it was mentioned at the end of section \ref{sec:Algorithm}, the constant $C_{\text{alg}}$ in a single core is sufficiently small to give accurate calculations up to $\hat{\nu}\sim 10^{-5}$ $\text{m}^{-1}$. We have plotted in figure \ref{subfig:MONKES_Scaling_Nfs} the cubic fit $C_{\text{alg}}N_{\text{fs}}^3$, where $C_{\text{alg}}=0.61(1/2025)^3\sim 7\cdot 10^{-11}$ s. 

As the {\texttt{LAPACK}} library is multithreaded and allows to parallelize the linear algebra operations through several cores, the scaling of {\MONKES} when running in parallel is represented. Additionally, for the resolutions selected in subsection \ref{subsec:Convergence}, we display in table \ref{tab:MONKES_Times_Multicore} the wall-clock time when running {\MONKES} using several cores in parallel. Note that for the W7-X cases, which require a smaller value of $N_{\text{fs}}$, the speed-up stalls at 8 cores. For CIEMAT-QI, that requires discretizing the flux-surface on a finer mesh, this does not happen in the range of cores considered.

\begin{table}[h]
	\centering
	\begin{tabular}{lccccc}
		\toprule
		\backslashbox{Case}{No. cores}   & 1  & 2  &  4 &  8 & 16\\ \midrule
		W7X-EIM $\widehat{E}_r=0$        & 22    & 13  &  8   &  5 &  5 \\
		W7X-EIM $\widehat{E}_r\ne 0$   & 40    & 20  & 12    & 8 &  6 \\ 
		W7X-KJM $\widehat{E}_r=0$        & 33    & 17  &  12 & 7 &   7   \\
		W7X-KJM $\widehat{E}_r\ne 0$   & 46    & 17  &  13 & 7 & 7  \\
		CIEMAT-QI $\widehat{E}_r=0$      & 78    & 45 & 29 & 21 & 16 \\
		CIEMAT-QI $\widehat{E}_r\ne 0$ & 78    & 45 & 29 & 21 & 16 \\\bottomrule
	\end{tabular}
	\caption{Wall-clock time of {\MONKES} in seconds for the selected triplets $(N_\theta,N_\zeta,N_\xi)$ when running in several cores.}
	\label{tab:MONKES_Times_Multicore}
\end{table}

\FloatBarrier 

\subsection{Benchmark of the monoenergetic coefficients}\label{subsec:Benchmark}
\begin{figure*}[h]	
	\captionsetup[sub]{skip=-1.75pt, margin=95pt}
	\centering
	%
	\tikzsetnextfilename{Benchmark-W7X-EIM-s0200-D11}	
	\begin{subfigure}[t]{0.32\textwidth}	
		\includegraphics{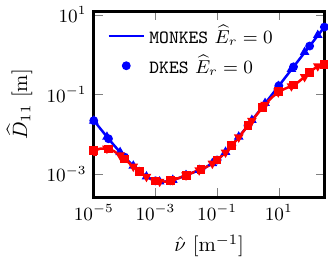}
		\caption{}
		\label{subfig:D11_Benchmark_W7X_EIM}
	\end{subfigure}
	\tikzsetnextfilename{Benchmark-W7X-KJM-s0204-D11}
	\begin{subfigure}[t]{0.32\textwidth}	
		\includegraphics{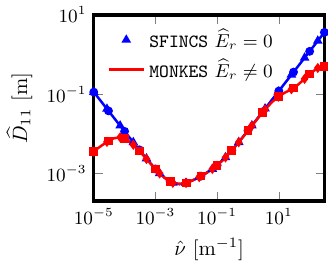}	
		\caption{}
		\label{subfig:D11_Benchmark_W7X_KJM}
	\end{subfigure}
	\tikzsetnextfilename{Benchmark-CIEMAT-QI-s0250-D11}
	\begin{subfigure}[t]{0.32\textwidth}	
		\includegraphics{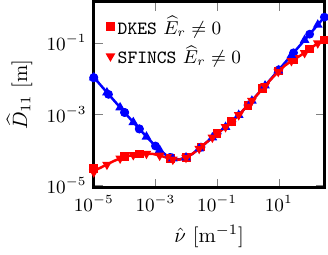}
		\caption{}
		\label{subfig:D11_Benchmark_CIEMAT_QI}
	\end{subfigure}
	\caption{Calculation of $\widehat{D}_{11}$ by \texttt{MONKES}, \texttt{DKES} and \texttt{SFINCS} for zero and finite $\widehat{E}_r$. (a) W7-X EIM at the surface $\psi /\psi_{\text{lcfs}}=0.200$. (b) W7-X KJM at the surface $\psi /\psi_{\text{lcfs}}=0.204$. (c) CIEMAT-QI at the surface $\psi /\psi_{\text{lcfs}}=0.250$. }
	\label{fig:D11_Benchmark}
\end{figure*}

\begin{figure*}[h]
	\centering
	\captionsetup[sub]{skip=-1.75pt, margin=80pt}
	\tikzsetnextfilename{Benchmark-W7X-EIM-s0200-D31}
	\begin{subfigure}[t]{0.32\textwidth}	
		\includegraphics{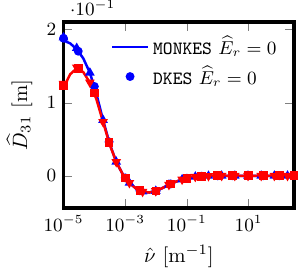}
		\caption{}
		\label{subfig:D31_Benchmark_W7X_EIM}
	\end{subfigure}
	\tikzsetnextfilename{Benchmark-W7X-KJM-s0204-D31}
	\begin{subfigure}[t]{0.32\textwidth}		
		\includegraphics{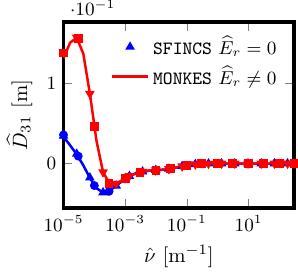}
		\caption{}
		\label{subfig:D31_Benchmark_W7X_KJM}
	\end{subfigure}
	\tikzsetnextfilename{Benchmark-CIEMAT-QI-s0250-D31}
	\begin{subfigure}[t]{0.32\textwidth}		
		\includegraphics{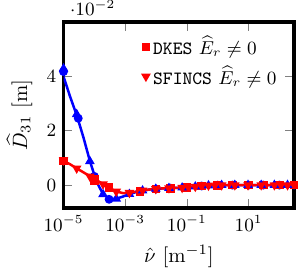}
		\caption{}
		\label{subfig:D31_Benchmark_CIEMAT_QI}
	\end{subfigure}
	\caption{Calculation of $\widehat{D}_{31}$ by \texttt{MONKES}, \texttt{DKES} and \texttt{SFINCS} for zero and finite $\widehat{E}_r$. (a) W7-X EIM at the surface $\psi /\psi_{\text{lcfs}}=0.200$. (b) W7-X KJM at the surface $\psi /\psi_{\text{lcfs}}=0.204$. (c) CIEMAT-QI at the surface $\psi /\psi_{\text{lcfs}}=0.250$. }
	\label{fig:D31_Benchmark}
\end{figure*}

\begin{figure*}[h]
	\captionsetup[sub]{skip=-1.75pt, margin=95pt}
	\centering
	%
	\tikzsetnextfilename{Benchmark-W7X-EIM-s0200-D33}
	\begin{subfigure}[t]{0.32\textwidth}	
		\includegraphics{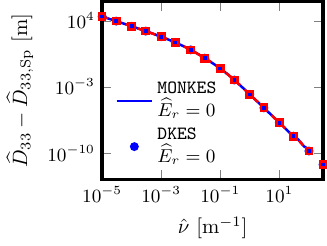}
		\caption{}
		\label{subfig:D33_Benchmark_W7X_EIM}
	\end{subfigure}
	\tikzsetnextfilename{Benchmark-W7X-KJM-s0204-D33}
	\begin{subfigure}[t]{0.32\textwidth}		
		\includegraphics{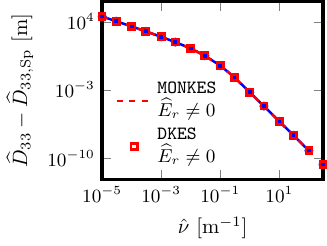}
		\caption{}
		\label{subfig:D33_Benchmark_W7X_KJM}
	\end{subfigure}
	\tikzsetnextfilename{Benchmark-CIEMAT-QI-s0250-D33}
	\begin{subfigure}[t]{0.32\textwidth}
		\includegraphics{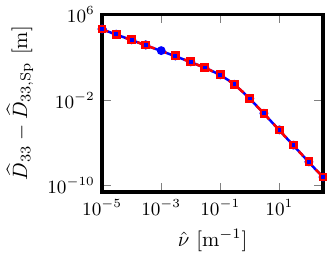}
		\caption{}
		\label{subfig:D33_Benchmark_CIEMAT_QI}
	\end{subfigure}
	\caption{Calculation of $\widehat{D}_{33}$ by \texttt{MONKES} and \texttt{DKES} for zero and finite $\widehat{E}_r$. (a) W7-X EIM at the surface $\psi /\psi_{\text{lcfs}}=0.200$. (b) W7-X KJM at the surface $\psi /\psi_{\text{lcfs}}=0.204$. (c) CIEMAT-QI at the surface $\psi /\psi_{\text{lcfs}}=0.250$.}
	\label{fig:D33_Benchmark}
\end{figure*}

Once we have chosen the resolutions $(N_\theta,N_\zeta,N_\xi)$ for each case, we need to verify that these selections indeed provide sufficiently accurate calculations of all the monoenergetic coefficients in the interval $\hat{\nu}\in[10^{-5},300]$ $\text{m}^{-1} $. In all cases, {\MONKES} calculations of the $\widehat{D}_{11}$ and $\widehat{D}_{31}$ coefficients will be benchmarked against converged calculations from {\DKES} (see \ref{sec:Appendix_DKES_Bounds}) and from \texttt{SFINCS}\footnote{\texttt{SFINCS} calculations are converged up to 3\% in the three independent variables.}. The parallel conductivity coefficient will be benchmarked only against {\DKES}. The benchmarking of the coefficient $\widehat{D}_{11}$ for the six different cases is shown in figure \ref{fig:D11_Benchmark}. The result of the benchmark of the bootstrap current coefficient $\widehat{D}_{31}$ is shown in figure \ref{fig:D31_Benchmark}. Finally, the parallel conductivity coefficient $\widehat{D}_{33}$ is benchmarked in figure \ref{fig:D33_Benchmark}. Due to the weak effect of the radial electric field in the $\widehat{D}_{33}$ coefficient, the symbols for this plot have been changed. In all cases, the agreement between {\MONKES}, {\DKES} and \texttt{SFINCS} is almost perfect. Thus, we conclude that {\MONKES} calculations of the monoenergetic coefficients are not only fast, but also accurate. Additionally, we can evaluate the level of optimization of the three configurations considered by inspecting these plots. In figures \ref{subfig:D11_Benchmark_W7X_EIM} and \ref{subfig:D11_Benchmark_W7X_KJM} is shown that the W7X-EIM configuration has smaller radial transport coefficient than the W7X-KJM configuration. Figures \ref{subfig:D31_Benchmark_W7X_EIM} and \ref{subfig:D31_Benchmark_W7X_KJM} show that the smaller radial transport of the W7X-EIM  configuration comes at the expense of having larger bootstrap current coefficient. As shown in figures \ref{subfig:D11_Benchmark_CIEMAT_QI} and \ref{subfig:D31_Benchmark_CIEMAT_QI}, the optimized stellarator CIEMAT-QI manages to achieve levels of radial transport similar or smaller than the W7X-EIM configuration and a bootstrap current coefficient as low as the W7X-KJM configuration.


\section{Conclusions and future work}
\label{sec:Conclusions}
In this paper we have presented the new code {\MONKES}, which can provide fast and accurate calculations of the monoenergetic transport coefficients at low collisionality in a single core. By means of a thorough convergence study we have shown that it is possible to evaluate the monoenergetic coefficients in the $1/\nu$ and $\sqrt{\nu}$-$\nu$ regimes in approximately 1 minute. Besides, when there are sufficient computational resources available, the code can run even faster using several cores in parallel. A natural application is the inclusion of {\MONKES} in a stellarator optimization suite. {\MONKES} rapid calculations will allow direct optimization of the bootstrap current and radial transport from low collisionalities (typical of the $1/\nu$ and $\sqrt{\nu}$-$\nu$) to moderate collisionalities (typical of the plateau regime). The low collisionality regimes are important in reactor relevant scenarios while the plateau regime can be important close to the edge, where the plasma is cooler. Massive evaluation of configurations to study the parametric dependence of $\widehat{D}_{31}$ or other coefficients on specific quantities of the magnetic configuration can also be done. Another application is its inclusion in predictive transport frameworks, which require neoclassical calculations to determine the evolution of plasma profiles. The neoclassical quantities required for these simulations can be calculated using {\MONKES}.

Equation (\ref{eq:DKE_Original}), solved by {\MONKES}, includes a collision operator which does not preserve momentum. An important continuation of this work would be the implementation of momentum-correction techniques, such as the ones presented in \cite{Taguchi,Sugama-PENTA,Sugama2008,MaasbergMomentumCorrection}. As each calculation from {\MONKES} can be executed in a single core, the scan in $v$ (i.e. in $\hat{\nu}$) required to perform the integrals of the monoenergetic coefficients is parallelizable. Therefore, it seems possible for the near future to obtain fast calculations of neoclassical transport with a model collision operator that preserves momentum. With this minor extension, {\MONKES} could also be used for self-consistent optimization of magnetic fields in a similar manner to \cite{Landreman_SelfConsistent} for general geometry.

\section*{Acknowledgements}
This work has been carried out within the framework of the EUROfusion Consortium, funded by the European Union via the Euratom Research and Training Programme (Grant Agreement No 101052200 – EUROfusion). Views and opinions expressed are however those of the author(s) only and do not necessarily reflect those of the European Union or the European Commission. Neither the European Union nor the European Commission can be held responsible for them. This research was supported in part by grant PID2021-123175NB-I00, Ministerio de Ciencia e Innovación, Spain. ML was supported by the U.S. Department of Energy, Office of Science, Office of Fusion Energy Science, under award number DE-FG02-93ER54197. This work was supported by the U.S. Department of Energy under contract number DE-AC02-09CH11466. The United States Government retains a non-exclusive, paid-up, irrevocable, world-wide license to publish or reproduce the published form of this manuscript, or allow others to do so, for United States Government purposes. The authors would like to thank C. D. Beidler for useful suggestions on the usage of {\DKES}.

\FloatBarrier
\appendix
\section{Onsager symmetry}
\label{sec:Appendix_Onsager_symmetry}
In this appendix we will prove that the monoenergetic coefficients $\widehat{D}_{ij}$ defined by (\ref{eq:Monoenergetic_geometric_coefficients}) satisfy Onsager symmetry relations whenever there is no electric field $E_\psi=0$ or the magnetic field possesses stellarator symmetry. For this, we will prove a more general result involving linear equations defined in some domain (phase-space) $\mathcal{S}$. Suppose we have a space $\mathcal{F_S}$ of functions from $\mathcal{S}$ to $\mathbb{R}$ with inner product $\mean*{\cdot,\cdot}_{\mathcal{S}}$ and a set of linear equations
\begin{align}
	\mathcal{V} f_j - \mathcal{C}f_j = s_j,
	\label{eq:Onsager_symmetry_DKE}
\end{align}
for $j=1,2\ldots, N_{\text{e}} $ where $s_j\in\mathcal{F_S}$ and the linear operators $\mathcal{C}$ and $\mathcal{V}$ are respectively symmetric and antisymmetric with respect to $\mean*{\cdot,\cdot}_{\mathcal{S}}$. Namely,
\begin{align}
	\mean*{\mathcal{C}f,g}_{\mathcal{S}} & = \mean*{f,\mathcal{C}g}_{\mathcal{S}}, 
	\label{eq:Onsager_symmetry_Collision_Symmetry}\\
	\mean*{\mathcal{V}f,g}_{\mathcal{S}} & = -\mean*{f,\mathcal{V}g}_{\mathcal{S}}.
	\label{eq:Onsager_symmetry_Vlasov_Skew_Symmetry}
\end{align}

Now, we define the scalars 
\begin{align}
	\mathcal{D}_{ij} := \mean*{ s_i, f_j}_{\mathcal{S}}
	\label{eq:Onsager_symmetry_coefficients}
\end{align}
for $i,j=1,2\ldots, N_{\text{e}} $.

Additionally, we define a property $\mathcal{P}$ to be a map which associates to each $f\in\mathcal{F_S}$ a function $\mathcal{P} f \in\mathcal{F_S}$ and is idempotent\footnote{This means that, for all $f\in\mathcal{F_S}$, $\mathcal{P} \mathcal{P} f=f$.}. Any function $f\in\mathcal{F_S}$ can be splitted in its even $f^+$ and odd $f^-$ portions with respect to the property $\mathcal{P}$ as follows
\begin{align}
	f^\pm : = 
	\frac{1}{2}
	\left(
	f \pm \mathcal{P} f
	\right),
\end{align}
satisfying $\mathcal{P} f^\pm = \pm f^\pm$. Without loss of generality, we assume that $N^+\le N_{\text{e}}$ sources $s_j$ in (\ref{eq:Onsager_symmetry_DKE}) are even with respect to $\mathcal{P}$ and the remaining $N^- := N_{\text{e}}- N^+$ sources are odd. 

The coefficients $\mathcal{D}_{ij}$ satisfy Onsager symmetry relations if three (sufficient) conditions are satisfied. 
\begin{enumerate}
	\item Even and odd functions are mutually orthogonal $\mean*{f^\pm, g^\mp}_\mathcal{S}=0$. This implies that
	\begin{align}
		\mean*{f,g}_{\mathcal{S}} = \mean*{f^+,g^+}_{\mathcal{S}} + \mean*{f^-,g^-}_{\mathcal{S}}.
		\label{eq:Onsager_symmetry_orthogonality_Even_Odd}
	\end{align}
	
	\item The operator $\mathcal{C}$ is even with respect to property $\mathcal{P}$. Explicitly,
	\begin{align}
		(\mathcal{C} f)^\pm & = \mathcal{C} f^\pm.
		\label{eq:Onsager_symmetry_Collisions_Even}
	\end{align}
	
	\item The operator $\mathcal{V}$ is odd with respect to property $\mathcal{P}$. Explicitly,
	\begin{align}
		(\mathcal{V} f)^\pm & = \mathcal{V} f^\mp.
		\label{eq:Onsager_symmetry_Vlasov_Odd}
	\end{align}
\end{enumerate}
When conditions (\ref{eq:Onsager_symmetry_orthogonality_Even_Odd}), (\ref{eq:Onsager_symmetry_Collisions_Even}) and (\ref{eq:Onsager_symmetry_Vlasov_Odd}) are satisfied we have the following Onsager symmetry relations.
\begin{itemize}
	\item For fixed $i$ and $j$, if $s_i$ and $s_j$ are both even, $\mathcal{D}_{ij} = \mathcal{D}_{ji}$. The proof is as follows 
	\begin{align*}
		\mathcal{D}_{ij} & = \mean*{s_i^+, f_j^+}_{\mathcal{S}} \\
		& 
		= \mean*{\mathcal{V} f_i^-, f_j^+}_{\mathcal{S}}
		- \mean*{\mathcal{C} f_i^+, f_j^+}_{\mathcal{S}} \\
		& 
		= - \mean*{f_i^-, \mathcal{V} f_j^+}_{\mathcal{S}}
		- \mean*{\mathcal{C} f_i^+, f_j^+}_{\mathcal{S}} \\
		& 
		= - \mean*{f_i^-, \mathcal{C} f_j^-}_{\mathcal{S}}
		- \mean*{\mathcal{C} f_i^+, f_j^+}_{\mathcal{S}}\\
		& 
		= - \mean*{f_i, \mathcal{C} f_j}_{\mathcal{S}}.
	\end{align*}
	As in the last equality, due to (\ref{eq:Onsager_symmetry_Collision_Symmetry}), the roles of $i$ and $j$ are interchangeable, we have that $\mathcal{D}_{ij} = \mathcal{D}_{ji}$.
	
	\item For fixed $i$ and $j$, if $s_i$ and $s_j$ are both odd, $\mathcal{D}_{ij} = \mathcal{D}_{ji}$. The proof is as follows 
	\begin{align*}
		\mathcal{D}_{ij} & = \mean*{s_i^-, f_j^-}_{\mathcal{S}} \\
		& 
		= \mean*{\mathcal{V} f_i^+, f_j^-}_{\mathcal{S}}
		- \mean*{\mathcal{C} f_i^-, f_j^-}_{\mathcal{S}} \\
		& 
		= - \mean*{f_i^+, \mathcal{V} f_j^-}_{\mathcal{S}}
		- \mean*{\mathcal{C} f_i^-, f_j^-}_{\mathcal{S}} \\
		& 
		= - \mean*{f_i^+, \mathcal{C} f_j^+}_{\mathcal{S}}
		- \mean*{\mathcal{C} f_i^-, f_j^-}_{\mathcal{S}} \\
		& 
		= - \mean*{f_i, \mathcal{C} f_j}_{\mathcal{S}}.
	\end{align*}
	As in the last equality, due to (\ref{eq:Onsager_symmetry_Collision_Symmetry}), the roles of $i$ and $j$ are interchangeable, we have that $\mathcal{D}_{ij} = \mathcal{D}_{ji}$.
	
	\item For fixed $i$ and $j$, if $s_i$ is even and $s_j$ is odd, $\mathcal{D}_{ij} = -\mathcal{D}_{ji}$. The proof is as follows 
	\begin{align*}
		\mathcal{D}_{ij} & = \mean*{s_i^+, f_j^+}_{\mathcal{S}} \\
		& 
		= \mean*{\mathcal{V} f_i^-, f_j^+}_{\mathcal{S}}
		- \mean*{\mathcal{C} f_i^+, f_j^+}_{\mathcal{S}} \\
		& 
		= \mean*{\mathcal{V} f_i^-, f_j^+}_{\mathcal{S}}
		- \mean*{f_i^+, \mathcal{C}  f_j^+}_{\mathcal{S}} \\
		& 
		= \mean*{\mathcal{V} f_i^-, f_j^+}_{\mathcal{S}}
		- \mean*{f_i^+, \mathcal{V}  f_j^-}_{\mathcal{S}} \\
		& 
		= \mean*{\mathcal{V} f_i, f_j}_{\mathcal{S}}.
	\end{align*}
	As in the last equality, due to (\ref{eq:Onsager_symmetry_Vlasov_Skew_Symmetry}), interchanging the roles of $i$ and $j$ switches signs, we have that $\mathcal{D}_{ij} = -\mathcal{D}_{ji}$.
	
\end{itemize}

The equation (\ref{eq:DKE}) can be written in the form of (\ref{eq:Onsager_symmetry_DKE}) by setting
the operators to be
\begin{align}
	\mathcal{V} & = \xi \vb*{b}\cdot \nabla + \nabla\cdot\vb*{b} \frac{1-\xi^2}{2}\pdv{\xi}
	-  \frac{\widehat{E}_\psi}{\mean*{B^2}} \vb*{B}\times \nabla \psi \cdot \nabla,
	\\ 
	\mathcal{C} & = \hat{\nu} \Lorentz,
\end{align}
and the inner product
\begin{align}
	\mean*{f,g}_{\mathcal{S}}:= \mean*{\int_{-1}^{1}fg\dd{\xi}}.
	\label{eq:Onsager_Inner_Product}
\end{align}

With these definitions, properties (\ref{eq:Onsager_symmetry_Collision_Symmetry}) and (\ref{eq:Onsager_symmetry_Vlasov_Skew_Symmetry})\footnote{As $\nabla \cdot \vb*{v}_E =0$, the operator $\mathcal{V}$ can be written in divergence form. For the symmetry of $\Lorentz$ see \ref{sec:Appendix_Legendre}.} are satisfied and $\mathcal{D}_{ij} = \widehat{D}_{ij} $. It is interesting to remark that the antisymmetry property (\ref{eq:Onsager_symmetry_Vlasov_Skew_Symmetry}) of $\mathcal{V}$ implies that the diagonal monoenergetic coefficients $\widehat{D}_{ii} $ are always positive. Note first that (\ref{eq:Onsager_symmetry_Vlasov_Skew_Symmetry}) implies $\mean*{f, \mathcal{V} f}_{\mathcal{S}} = 0$ for any $f\in \mathcal{F}_{\mathcal{S}}$. This implies that $\widehat{D}_{ii} = - \mean*{f_i, \hat{\nu}\Lorentz f_i}_{\mathcal{S}} $ and, as $\Lorentz$ is a negative operator (its eigenvalues are all negative or zero, see \ref{sec:Appendix_Legendre}), $\widehat{D}_{ii} \ge 0$. Also note that properties (\ref{eq:Onsager_symmetry_Collision_Symmetry}) and (\ref{eq:Onsager_symmetry_Collision_Symmetry}) imply that $\mean*{\hat{\nu}\mathcal{L}f_j,1}_{\mathcal{S}} = 0$ and $\mean*{\mathcal{V}f_j,1}_{\mathcal{S}} = 0$. Thus, the image of the drift-kinetic equation (\ref{eq:DKE}) is constrained by $\mean*{s_j,1}_{\mathcal{S}} = 0$.

Now we distinguish the two cases for which the monoenergetic coefficients $\widehat{D}_{ij} $ satisfy Onsager symmetry relations. Apart from the velocity coordinate $\xi$, we will use Boozer coordinates $( {\theta},\zeta)$.
\begin{enumerate}
	\item If $E_\psi=0$, the property is defined as 
	\begin{align}
		\mathcal{P} f( {\theta},\zeta,\xi) = f( {\theta},\zeta,-\xi).
	\end{align}
	It is straightforward to check that for this property, conditions (\ref{eq:Onsager_symmetry_orthogonality_Even_Odd}), (\ref{eq:Onsager_symmetry_Collisions_Even}) and (\ref{eq:Onsager_symmetry_Vlasov_Odd}) are satisfied. Also, $s_1= s_1^+$, $s_2 = s_2^+$ and $s_3 = s_3^-$. Hence, we have $\widehat{D}_{12}=\widehat{D}_{21}$, $\widehat{D}_{13}=-\widehat{D}_{31}$ and $\widehat{D}_{23}=-\widehat{D}_{32}$.

	\item When $E_\psi$ is not necessarily zero, we define the property $\mathcal{P}$ as the one that defines stellarator symmetry \cite{DEWAR1998275}
	\begin{align}
		\mathcal{P} f( {\theta},\zeta,\xi) = f( - {\theta}, - \zeta,\xi)
	\end{align}
	and we have assumed without loss of generality that the planes of symmetry are $\theta=0$ and $ \zeta =0$. Thus, when the magnetic field is stellarator-symmetric $B=B^+$. In this case, using (\ref{eq:FSA_Boozer}), (\ref{eq:Parallel_streaming_spatial_operator}) and (\ref{eq:ExB_spatial_operator}) it is straightforward to check\footnote{Note that derivatives along $\theta$ and $\zeta$ switch parities with respect to the stellarator symmetry property, i.e. $\pdv*{f^\pm}{\theta} = (\pdv*{f^\pm}{\theta} )^\mp$ and $\pdv*{f^\pm}{\zeta} = (\pdv*{f^\pm}{\zeta} )^\mp$. Also, as for stellarator-symmetric fields, $\sqrt{g}=\sqrt{g}^+$ the flux-surface average satisfies $\mean*{f^-}=0$.} that conditions (\ref{eq:Onsager_symmetry_orthogonality_Even_Odd}), (\ref{eq:Onsager_symmetry_Collisions_Even}) and (\ref{eq:Onsager_symmetry_Vlasov_Odd}) are satisfied. Besides, $s_1=s_1^-$, $s_2 = s_2^-$ and $s_3 = s_3^+$. Hence, we have $\widehat{D}_{12}=\widehat{D}_{21}$, $\widehat{D}_{13}=-\widehat{D}_{31}$ and $\widehat{D}_{23}=-\widehat{D}_{32}$. 
\end{enumerate}

Note that for equation (\ref{eq:DKE}), the Onsager symmetry relation $\widehat{D}_{12}=\widehat{D}_{21}$ is trivial as $s_1=s_2$, which implies $f_1=f_2$ and thus $\widehat{D}_{12}=\widehat{D}_{21} = \widehat{D}_{11}= \widehat{D}_{22} $, $\widehat{D}_{31} = \widehat{D}_{32}$ and $\widehat{D}_{13} = \widehat{D}_{23}$. Nevertheless, if the definition of $s_1$ and $s_2$ was different, as long as their parity is the same, the relation $\widehat{D}_{12}=\widehat{D}_{21}$ would still hold.

\section{Legendre modes of the drift-kinetic equation}\label{sec:Appendix_Legendre}

Legendre polynomials are the eigenfunctions of the Sturm-Liouville problem in the interval $\xi\in[-1,1]$ defined by the differential equation%
\begin{align}
	2\Lorentz P_k(\xi) = -k(k+1) P_k(\xi), \label{eq:Legendre_eigenvalues}
\end{align}
and regularity boundary conditions at $\xi = \pm 1 $
\begin{align}
	\eval{(1-\xi^2)\dv{P_k}{\xi}}_{\xi = \pm 1} = 0,
\end{align}
where $k\ge 0$ is an integer. 

As $\Lorentz$ has a discrete spectrum and is self-adjoint with respect to the inner product
\begin{align}
	\mean*{f,g}_\Lorentz := \int_{-1}^{1} fg \dd{\xi} ,
\end{align}
in the space of functions that satisfy the regularity condition, $\{P_k\}_{k=0}^{\infty}$ is an orthogonal basis satisfying $\mean*{P_j,P_k}_\Lorentz = 2 \delta_{jk}/(2k+1)$. Hence, these polynomials satisfy the three-term recurrence formula
\begin{align}
	(2k+1)\xi P_k(\xi) = (k+1)P_{k+1}(\xi) + k P_{k-1}(\xi),
	\label{eq:Legendre_Three_Term_Recurrence}
\end{align}
obtained by Gram-Schmidt orthogonalization. Starting from the initial values $P_0=1$ and $P_1=\xi$, the recurrence defines the rest of the Legendre polynomials. Additionally, they satisfy the differential identity
\begin{align}
	(1-\xi^2)\dv{P_k}{\xi} = k P_{k-1}(\xi) - k \xi P_k(\xi).
	\label{eq:Legendre_Differential_Recurrence}
\end{align}
Identities (\ref{eq:Legendre_Three_Term_Recurrence}) and (\ref{eq:Legendre_Differential_Recurrence}) are useful to represent tridiagonally the left-hand side of equation (\ref{eq:DKE}) when we use the expansion (\ref{eq:Legendre_expansion}). The $k-$th Legendre mode of the term $\xi\vb*{b}\cdot \nabla f $ is expressed in terms of the modes $f^{(k-1)}$ and $f^{(k+1)}$ using (\ref{eq:Legendre_Three_Term_Recurrence})
\begin{align}	
	\mean*{\xi \vb*{b}\cdot\nabla f, P_k}_\Lorentz
	=
	\frac{2}{2k+1}
	\left[
	\frac{k}{2k-1} 
	\right.
	& \vb*{b}\cdot\nabla f^{(k-1)} \nonumber
	\\+
	\frac{k+1}{2k+3} 
	& 
	\left.\vb*{b}\cdot\nabla f^{(k+1)} 
	\right].
\end{align}
Combining both (\ref{eq:Legendre_Three_Term_Recurrence}) and (\ref{eq:Legendre_Differential_Recurrence}) allows to express the $k-$th Legendre mode of the mirror term $\nabla\cdot \vb*{b}((1-\xi^2)/2)\pdv*{f}{\xi}$ in terms of the modes $f^{(k-1)}$ and $f^{(k+1)}$ as
\begin{align} 	
	& \mean*{ 
		\frac{1}{2}(1-\xi^2)
		\nabla\cdot\vb*{b}  
		\pdv{f}{\xi}, P_k}_\Lorentz
	=	\\
	&\frac{\vb*{b}\cdot\nabla \ln B}{2k+1}
	\left[
	\frac{k(k-1)}{2k-1} 
	f^{(k-1)} 
	-
	\frac{(k+1)(k+2)}{2k+3}
	f^{(k+1)} 
	\right], \nonumber
\end{align}
where we have also used $\nabla\cdot\vb*{b}  = - \vb*{b}\cdot \nabla \ln B$. The term proportional to $\widehat{E}_\psi$ is diagonal in a Legendre representation
\begin{align}
	&
	\mean*{\frac{\widehat{E}_\psi}{\mean*{B^2}}
		\vb*{B}\times \nabla\psi \cdot\nabla f , P_k}_\Lorentz
	=
	\\ 
	&
	\frac{2}{2k+1}
	\frac{\widehat{E}_\psi}{\mean*{B^2}}
	\vb*{B}\times \nabla\psi \cdot\nabla f^{(k)}.
	\nonumber
\end{align}
For the collision operator used in equation (\ref{eq:DKE}), as Legendre polynomials are eigenfunctions of the pitch-angle scattering operator, using (\ref{eq:Legendre_eigenvalues}) we obtain the diagonal representation 
\begin{align}
	\mean*{\hat{\nu} \Lorentz f , P_k}_\Lorentz
	&
	=
	-\hat{\nu}
	\frac{k(k+1)}{2k+1}	
	f^{(k)}.
\end{align}

Finally, we briefly comment on why the truncation error from (\ref{eq:Legendre_expansion}) implies that the solution to (\ref{eq:DKE_Legendre_expansion}) and (\ref{eq:kernel_elimination_condition_Legendre}) is an approximation of the Legendre spectrum of the exact solution to (\ref{eq:DKE}) satisfying (\ref{eq:kernel_elimination_condition}). For this, we will assume that the solution to (\ref{eq:DKE}) and (\ref{eq:kernel_elimination_condition}) is unique (which it is, see \ref{sec:Appendix_Invertibility}). We denote this exact solution by $f_{\text{ex}}$ and its Legendre modes by $f^{(k)}_{\text{ex}}$. The Legendre modes $f^{(k)}_{\text{ex}}$ satisfy (\ref{eq:DKE_Legendre_expansion}) for all values of $k$, including $k>N_\xi$ and, in general, $f^{(N_\xi+1)}_{\text{ex}}\ne 0$. Denoting the error of the solution $f^{(k)}$ to (\ref{eq:DKE_Legendre_expansion}) and (\ref{eq:kernel_elimination_condition_Legendre}) by
\begin{align}
	E^{(k)} : = f^{(k)}_{\text{ex}} - f^{(k)} ,
\end{align}
is easy to prove that
\begin{align}
	L_k E^{(k-1)} + D_k E^{(k)} + U_k E^{(k+1)} = 0, 
	\label{eq:Error_DKE_Legendre_k}
\end{align}
for $k =0,1,\ldots, N_\xi-1$ and 
\begin{align}
	L_{N_\xi} E^{(N_\xi-1)} + D_{N_\xi} E^{(N_\xi)} = - U_{N_\xi} f^{(N_\xi+1)}_{\text{ex}}.
	\label{eq:Error_DKE_Legendre_Nxi}
\end{align}
Note that the system of equations constituted by (\ref{eq:Error_DKE_Legendre_k}) and (\ref{eq:Error_DKE_Legendre_Nxi}) for the error is identical to (\ref{eq:DKE_Legendre_expansion}) substituting $f^{(k)}$ by $E^{(k)}$ and $s^{(k)}$ by $- U_{N_\xi} f^{(N_\xi+1)}_{\text{ex}}$. Hence, by assumption, the solution to (\ref{eq:Error_DKE_Legendre_k}) and (\ref{eq:Error_DKE_Legendre_Nxi}) satisfying (\ref{eq:kernel_elimination_condition_Legendre}) is unique, implying that $E^{(k)}\ne 0$ unless $ {U}_{N_\xi} f^{(N_\xi+1)}_{\text{ex}} = 0$.

\section{Invertibility of the spatial differential operators}
\label{sec:Appendix_Invertibility} 
In this Appendix we will study
the invertibility of the left-hand-side of (\ref{eq:DKE_Legendre_expansion}). We are only concerned in elucidating under which conditions the algorithm given in section \ref{sec:Algorithm} can be applied to solve (\ref{eq:DKE_Legendre_expansion}). For instance, we will consider the possibility of the flux-surface being rational despite of the fact that (among other things) it may be inconsistent with the assumption that thermodynamical forces are a flux-function. We will conclude that the solution to (\ref{eq:DKE_Legendre_expansion}) submitted to (\ref{eq:kernel_elimination_condition_Legendre}) is unique in ergodic flux-surfaces and also on rational flux-surfaces with $E_\psi\ne 0$ and can be obtained with the aforementioned algorithm. In order to do this, we view $L_k$, $D_k$ and $U_k$ as operators that act on $\mathcal{F}$, where $\mathcal{F}$ is the space of smooth functions on the flux-surface equipped with the inner product
\begin{align}
	\mean*{f,g}_{\mathcal{F}}=\frac{N_p}{4\pi^2}\oint\oint f\bar{g}\dd{\theta}\dd{\zeta},
	\label{eq:Fourier_inner_product}
\end{align}
where $\bar{z}$ denotes the complex conjugate of $z$ and the inner product induces a norm 
\begin{align}
	\norm{f}_{\mathcal{F}}:=\sqrt{\mean*{f,f}_{\mathcal{F}}}.
	\label{eq:Fourier_norm}
\end{align}
In this setting $L_k$, $D_k$ and $U_k$ are operators from $\mathcal{F}$ to $\mathcal{F}$ as all of their coefficients are smooth on the flux-surface. However, the operators $L_k$ and $U_k$ given by (\ref{eq:DKE_Legendre_expansion_Lower}) and (\ref{eq:DKE_Legendre_expansion_Upper}) do not have a uniquely defined inverse. This is a consequence of the fact that the parallel streaming operator $\xi \vb*{b}\cdot \nabla+\nabla \cdot \vb*{b} {(1-\xi^2)}/{2}  \pdv*{\xi}$ has a non trivial kernel comprised of functions $g((1-\xi^2)/B)$. On the other hand, the operator $D_k$ has a unique inverse for $k\ge 1$. For $k=0$, the operator $D_0$ is not invertible as it has a kernel comprised of functions $g(B_\theta\theta + B_\zeta\zeta)$.

Whether $L_k$ and $U_k$ are or not invertible can be determined studying the uniqueness of continuous solutions (on the flux-surface) to
\begin{align}
	\vb*{B}\cdot \nabla f + \omega_k f = s B,
	\label{eq:Invertibility_Lk_Uk_MDE}
\end{align}
for some $s,\omega_k\in\mathcal{F}$. Note that equations $L_k f = ks/(2k-1)$ and $U_k f =(k+1)s/(2k+3)$ can be written in the form of equation (\ref{eq:Invertibility_Lk_Uk_MDE}) setting, respectively, $\omega_k=(k-1)\vb*{B}\cdot\nabla \ln B /2$ and $\omega_k=-(k+2)\vb*{B}\cdot\nabla \ln B /2$. We will determine a condition for $\omega_k$ which, if satisfied, equation (\ref{eq:Invertibility_Lk_Uk_MDE}) has a unique solution $f\in\mathcal{F}$.

The solution to equation (\ref{eq:Invertibility_Lk_Uk_MDE}) can be written as
\begin{align}
	f = (f_0 + K ) \Phi,
	\label{eq:Invertibility_Lk_Uk_Variation_of Constants}
\end{align}
where
\begin{align}
	& \vb*{B}\cdot \nabla f_0 = 0,  \label{eq:Invertibility_Lk_Uk_constant}
	\\
	& \vb*{B}\cdot \nabla \Phi + \omega_k \Phi = 0,  \label{eq:Invertibility_Lk_Uk_Homogeneous}
	\\
	& \vb*{B}\cdot \nabla K = {sB}/{\Phi}.  \label{eq:Invertibility_Lk_Uk_Particular}
\end{align}
Equations (\ref{eq:Invertibility_Lk_Uk_Homogeneous}) and (\ref{eq:Invertibility_Lk_Uk_Particular}) are integrated (along a field line) imposing $\eval{\Phi}_{p}=1$ and $\eval{K}_{p}=0$ at a point $p$ of the field line. Note that $f_0=\eval{f}_{p}$ is an integration constant. Depending on the form of $\omega_k$, $f_0$ can or cannot be determined imposing continuity on the flux-surface. The solution to equation (\ref{eq:Invertibility_Lk_Uk_Homogeneous}) can be written as
\begin{align}
	\Phi = \exp(-W_k), 
	\label{eq:Invertibility_Lk_Uk_Homogeneous_solution}
\end{align}
where $\vb*{B}\cdot\nabla W_k = \omega_k$ and is integrated imposing $\eval{W_k}_{p}=0$. Note that this implies that $\Phi\ne 0$ and that 
\begin{align}
	- \vb*{B}\cdot \nabla \left(\frac{1}{\Phi}\right) + \omega_k \frac{1}{\Phi} = 0.
	\label{eq:Invertibility_Lk_Uk_Adjoint_kernel}
\end{align}
When $\Phi\in\mathcal{F}$, the left-hand side of (\ref{eq:Invertibility_Lk_Uk_MDE}) has a non trivial kernel (as an operator from $\mathcal{F}$ to $\mathcal{F}$). In order to proceed further, we employ coordinates $(\alpha,l)$ where $\alpha:=\theta-\iota \zeta$ is a poloidal angle that labels field lines and $l$ is the length along magnetic field lines. Depending on the type of flux-surface there are two possible situations
\begin{enumerate}
	\item For ergodic flux-surfaces, $\iota\in\mathbb{R}\backslash \mathbb{Q}$ and satisfying (\ref{eq:Invertibility_Lk_Uk_constant}) implies that $f_0$ is a flux-function. The solution $f$ to (\ref{eq:Invertibility_Lk_Uk_MDE}) is a differentiable function on the torus if $\mean*{\vb*{B}\cdot\nabla f} = 0$. Applying $\mean*{\text{Eq. (\ref{eq:Invertibility_Lk_Uk_MDE})}}$ combined with splitting (\ref{eq:Invertibility_Lk_Uk_Variation_of Constants}) yields
	\begin{align}
		f_0\mean*{\omega_k \Phi}  & = \mean*{B s} - \mean*{K \omega_k \Phi} \nonumber
		\\
		& = \mean*{\vb*{B}\cdot\nabla(K \Phi)}. \label{eq:Invertibility_Lk_Uk_Ergodic_condition}
	\end{align}
	Hence, if $\mean*{\omega_k \Phi} \ne 0$, equation (\ref{eq:Invertibility_Lk_Uk_Ergodic_condition}) fixes the value of $f_0$ so that $f$ is continuous on the torus. Note that if $\mean*{\omega_k \Phi} \ne 0$, by virtue of (\ref{eq:Invertibility_Lk_Uk_Homogeneous}), $\Phi$ is not univaluated and does not belong to $\mathcal{F}$. On the contrary, if $f_0$ is free, then $\Phi$ is a continuous function on the torus. Then, (\ref{eq:Invertibility_Lk_Uk_Ergodic_condition}) implies that $K\Phi$ is continuous on the torus when $\Phi$ is. The function $K$ is also continuous as long as $sB$ belongs to the image of $\vb*{B}\cdot \nabla + \omega_k$. Note that using (\ref{eq:Invertibility_Lk_Uk_Adjoint_kernel}) we can derive from $\mean*{\text{Eq. (\ref{eq:Invertibility_Lk_Uk_MDE})}/\Phi}$ the solvability condition $\mean*{sB/\Phi}=0$.
	
	\item For rational flux-surfaces, $\iota\in \mathbb{Q}$ and satisfying (\ref{eq:Invertibility_Lk_Uk_constant}) implies that $f_0(\alpha)$ depends on the field line chosen. At these surfaces, the field line labelled by $\alpha$ closes on itself after a length $L_{\text{c}}(\alpha)$. If the solution $f$ is continuous on the flux-surface, then $\int_{0}^{L_{\text{c}}} \vb*{B}\cdot\nabla f \dd{l}/B=0$ for each field line. Applying $\int_{0}^{L_{\text{c}}} \text{Eq. (\ref{eq:Invertibility_Lk_Uk_MDE})}\dd{l}/B$ combined with splitting (\ref{eq:Invertibility_Lk_Uk_Variation_of Constants}) yields
	\begin{align}
		f_0(\alpha)
		\int_{0}^{L_{\text{c}}} \omega_k \Phi \frac{\dd{l}}{B}  
		& =  
		\int_{0}^{L_{\text{c}}} s \dd{l} - \int_{0}^{L_{\text{c}}} \omega_k K \Phi \frac{\dd{l}}{B}
		\nonumber 
		\\
		& =  
		\int_{0}^{L_{\text{c}}} \vb*{B}\cdot\nabla(K \Phi) \frac{\dd{l}}{B}
		. \label{eq:Invertibility_Lk_Uk_Rational_condition}
	\end{align}
	If $\int_{0}^{L_{\text{c}}} \omega_k \Phi {\dd{l}}/{B} \ne 0$, condition (\ref{eq:Invertibility_Lk_Uk_Rational_condition}) fixes a unique value of $f_0(\alpha)$ (for each field line) for which $f$ is continuous on the torus. As for ergodic surfaces, if (\ref{eq:Invertibility_Lk_Uk_Rational_condition}) does not fix $f_0$, then $\Phi$ and $K\Phi$ are continuous along field lines. Again, $K$ is also continuous as long as $sB$ belongs to the image of $\vb*{B}\cdot \nabla + \omega_k$. Using (\ref{eq:Invertibility_Lk_Uk_Adjoint_kernel}) we can derive from $\int_{0}^{L_{\text{c}}}\text{Eq. (\ref{eq:Invertibility_Lk_Uk_MDE})}/\Phi{\dd{l}}/{B}$ the solvability condition $\int_{0}^{L_{\text{c}}}sB/\Phi{\dd{l}}/{B}=0$.
\end{enumerate}

Thus, we have seen that when $\mean*{\omega_k \Phi}=0$ or $\int_{0}^{L_{\text{c}}} \omega_k \Phi {\dd{l}}/{B}=0$, the operator $\vb*{B}\cdot\nabla + \omega_k$ from $\mathcal{F}$ to itself is not one-to-one (it has a non trivial kernel comprised of multiples of $\Phi$).  
Moreover, we have the solvability conditions $\mean*{sB/\Phi}=0$ for ergodic surfaces and $\int_{0}^{L_\text{c}}sB/\Phi\dd{l}/B = 0$ for rational surfaces. The existence of a solvability condition implies that $\vb*{B}\cdot\nabla + \omega_k$ is not onto. We can derive a simpler and equivalent condition for $\omega_k$ from (\ref{eq:Invertibility_Lk_Uk_Homogeneous_solution}). Note that $\Phi$ is continuous on the torus only when $W_k$ is. As $\vb*{B}\cdot\nabla W_k = \omega_k$, continuity of $W_k$ along field lines imposes $\mean*{\omega_k}=0$ on ergodic flux-surfaces and $\int_{0}^{L_{\text{c}}} \omega_k  {\dd{l}}/{B} =0$ on rational ones. Hence, the operator $\vb*{B}\cdot\nabla + \omega_k$ is invertible if $\mean*{\omega_k} \ne 0 $ or $\int_{0}^{L_{\text{c}}} \omega_k {\dd{l}}/{B} \ne 0$. 

This result can be applied to determine that $L_k$ and $U_k$ are not invertible. For both $L_k$ and $U_k$, $\omega_k \propto \vb*{B}\cdot\nabla\ln B^\gamma$ for some rational exponent $\gamma$. As $B$ is continuous on the flux-surface we have for $L_k$ and $U_k$ that $\int_{0}^{L_{\text{c}}} \omega_k  {\dd{l}}/{B} =  0$ or $\mean*{\omega_k }=0$, which means that neither $L_k$ nor $U_k$ are invertible. 

Now we turn our attention to the invertibility of $D_k$ for $k\ge 1$. For $\widehat{E}_\psi =0$, $D_k$ is just a multiplicative operator and is clearly invertible when $\hat{\nu}, k\ne0$. For $\widehat{E}_\psi \ne 0$, the invertibility of $D_k$ can be proven by studying the uniqueness of solutions to
\begin{align} 
	\vb*{B}\times\nabla\psi\cdot \nabla g - \hat{\nu}_k g  = - \frac{\mean*{B^2}}{\widehat{E}_\psi} s,
	\label{eq:Invertibility_Dk_MDE}
\end{align}
where $\hat{\nu}_k = \hat{\nu} k (k+1) {\mean*{B^2}}/{2\widehat{E}_\psi}$. The procedure is very similar to the one carried out for $L_k$ and $U_k$. First, we write the solution to equation (\ref{eq:Invertibility_Dk_MDE}) as
\begin{align}
	g = ( g_0 + I ) \Psi,
	\label{eq:Invertibility_Dk_Variation_of_Constants}
\end{align}
where
\begin{align}
	& \vb*{B}\times\nabla\psi\cdot \nabla g_0 = 0,  \label{eq:Invertibility_Dk_constant}
	\\
	& \vb*{B}\times\nabla\psi\cdot \nabla \Psi - \hat{\nu}_k \Psi = 0,  \label{eq:Invertibility_Dk_Homogeneous}
	\\
	& \vb*{B}\times\nabla\psi\cdot \nabla I = - \frac{\mean*{B^2}}{\widehat{E}_\psi}\frac{s}{\Psi}.  \label{eq:Invertibility_Dk_Particular}
\end{align}
Equations (\ref{eq:Invertibility_Dk_Homogeneous}) and (\ref{eq:Invertibility_Dk_Particular}) are integrated along a integral curve of $\vb*{B}\times\nabla\psi$ imposing $\eval{\Psi}_{p}=1$ and $\eval{I}_{p}=0$ at the initial point $p$ of integration. The integral curves of $\vb*{B}\times\nabla\psi$ are, in Boozer coordinates, straight lines $B_\theta \theta + B_\zeta \zeta = \text{constant}$. In order to proceed further, we change from Boozer angles $(\theta,\zeta)$ to a different set of magnetic coordinates $(\alpha,\varphi)$ using the linear transformation
\begin{align}
	\Matrix{c}{\theta \\ \zeta}
	=
	\Matrix{cc}
	{
		(1 + \iota \delta)^{-1}	 &  \iota\\ 
		-\delta(1 + \iota \delta)^{-1}	 &  1
	}
	\Matrix{c}{\alpha \\ \varphi}
	\label{eq:Invertibility_Magnetic_Coordinates_Dk}
\end{align}
where $\delta = B_\theta/B_\zeta$. In these coordinates $\vb*{B} =\nabla \psi \times \nabla \alpha $, $B_\alpha = 0$ and
\begin{align}
	\vb*{B}\times\nabla \psi \cdot \nabla = 
	B^2 \pdv{}{\alpha}.
	\label{eq:Invertibility_Dk_Localization_alpha}
\end{align}
Depending on the rationality or irrationality of $\delta$ we can distinguish two options
\begin{enumerate}
	\item If $\delta \in \mathbb{R}\backslash\mathbb{Q}$, satisfying (\ref{eq:Invertibility_Dk_constant}) implies that $g_0$ is a flux-function (the integral curves trace out the whole flux-surface). Note that if $g$ is a differentiable function on the torus $\mean{ \vb*{B} \times \nabla \psi \cdot \nabla g } = \mean{ \nabla\times (g\vb*{B}) \cdot\nabla \psi }=0$, where we have used $\nabla\times\vb*{B}\cdot\nabla\psi = 0$. Taking $\mean*{\text{Eq. (\ref{eq:Invertibility_Dk_MDE})}}$ assuming that $f$ is continuous on the flux-surface, combined with (\ref{eq:Invertibility_Dk_Variation_of_Constants}) gives
	\begin{align} 
		\mean*{\Psi} g_0  
		& = 
		\frac{ \mean*{B^2} }{ \hat{\nu}_k \widehat{E}_\psi } 
		\mean*{s}
		- 
		\mean*{I \Psi}\nonumber
		\\
		& = 
		\frac{1}{\hat{\nu}_k}
		\mean*{\vb*{B}\times\nabla \psi \cdot \nabla(I \Psi)}. 
		\label{eq:Invertibility_Dk_Ergodic_condition}
	\end{align}
	Hence, if $\mean*{\Psi}\ne 0$, continuity of $g$ on the torus fixes the integration constant $g_0$. 
	
	\item If $\delta \in \mathbb{Q}$, satisfying (\ref{eq:Invertibility_Dk_constant}) implies that $g_0(\varphi)$ is a function of $\varphi$. Now the integral curves $\varphi=\text{constant}$ close on itself after moving in $\alpha$ an arc-length $L_\alpha$. In this scenario, if $g$ is a differentiable function on the torus $\int_{0}^{L_\alpha}\vb*{B}\times \nabla\psi \cdot \nabla g \dd{\alpha}/ B^2  = 0$, where we have used (\ref{eq:Invertibility_Dk_Localization_alpha}). Thus, taking $\int_{0}^{L_\alpha}\text{Eq. (\ref{eq:Invertibility_Dk_MDE})} \dd{\alpha}/ B^2 $, combined with (\ref{eq:Invertibility_Dk_Variation_of_Constants}) gives
	\begin{align}
		g_0(\varphi) \int_{0}^{L_\alpha} \Psi\frac{\dd{\alpha}}{ B^2 }
		& = 
		\frac{ \mean*{B^2} }{ \hat{\nu}_k \widehat{E}_\psi }
		\int_{0}^{L_\alpha} s    \frac{\dd{\alpha}}{ B^2 }
		\nonumber 
		-      
		\int_{0}^{L_\alpha}   I \Psi  \frac{\dd{\alpha}}{ B^2 }
		\\
		& =       
		\frac{1}{\hat{\nu}_k}
		\int_{0}^{L_\alpha}
		\vb*{B}\times\nabla \psi \cdot \nabla(I \Psi)
		\frac{\dd{\alpha}}{ B^2 }
		.
		\label{eq:Invertibility_Dk_Rational_condition}
	\end{align}
	Thus, if $\int_{0}^{L_\alpha} \Psi \dd{\alpha} / B^2 \ne 0 $ condition (\ref{eq:Invertibility_Dk_Rational_condition}) fixes the value of $g_0(\varphi)$ so that $g$ is continuous on the flux-surface. 
	
\end{enumerate}
Similarly to what happened to $\Phi$ when studying the invertibility of $L_k$ and $U_k$, continuity of the solution implies that $\Psi$ cannot be univaluated. We can write $\Psi$ as
\begin{align}
	\Psi = \exp(-A_k), 
	\label{eq:Invertibility_Dk_Exponential}
\end{align}
where $\vb*{B}\times \nabla\psi \cdot \nabla A_k = \hat{\nu}_k$ and is integrated along with condition $\eval{A_k}_{p}=0$. Using (\ref{eq:Invertibility_Dk_Localization_alpha}), we can write
\begin{align}
	A_k(\alpha,\varphi) = \hat{\nu}_k \int_{0}^{\alpha} \frac{\dd{\alpha'}}{ B^2(\alpha',\varphi)}.
\end{align}
Note that $A_k$ is monotonically crescent with $\alpha$, which means that $\Psi $ cannot be univaluated. Besides, (\ref{eq:Invertibility_Dk_Exponential}) implies $\Psi>0$, which means that $\mean*{\Psi}\ne 0 $ and $\int_{0}^{L_\alpha}\Psi   \dd{\alpha}/  B^2 \ne 0 $. Thus, there is a unique value of the constant $g_0$ which compensates the jumps in $\Psi$ and $I\Psi$ so that $g=g_0 \Psi + I \Psi$ is continuous on the flux-surface. Hence, $D_k$ is an invertible operator from $\mathcal{F}$ to itself.

The inverse of $D_k$ for $k \ge 1$ and $\widehat{E}_\psi \ne 0$ is defined by
\begin{align}
	D_k^{-1} s :=  ( \mathcal{G}_0[s] + \mathcal{I}[s] ) \Psi,
\end{align}
where $\mathcal{G}_0[s]$ and $\mathcal{I}[s]$ denote the linear operators which define, respectively, the constant of integration and the solution to (\ref{eq:Invertibility_Dk_Particular}) with $\eval{I}_{p}=0$ for a given source term. Specifically,
\begin{align}
	\mathcal{I}[s](\alpha,\varphi)
	& :=
	- \frac{\mean*{B^2}}{\widehat{E}_\psi}
	\int_{0}^{\alpha}
	\frac{s(\alpha',\varphi)}{\Psi(\alpha',\varphi)}
	\frac{\dd{\alpha'}}{ B^2(\alpha',\varphi)}  ,
\end{align}
and
\begin{align}
	\mathcal{G}_0[s](\varphi)
	& :=
	\begin{dcases}
		& \text{If }\delta\in\mathbb{R}\backslash\mathbb{Q}:\\
		& \frac{2}{ \hat{\nu} k (k+1) } \frac{\mean*{s}}{\mean*{\Psi} }
		- 
		\frac{\mean*{\mathcal{I}[s] \Psi}}{\mean*{\Psi} },  
		\vspace{0.5cm} \\ 
		& \text{If }\delta\in\mathbb{Q}:\\
		& \frac{2}{\hat{\nu} k(k+1)} \dfrac{\int_{0}^{L_\alpha} s \frac{\dd{\alpha}}{ B^2 }}{\int_{0}^{L_\alpha} \Psi\frac{\dd{\alpha}}{ B^2 }}
		-      
		\dfrac{\int_{0}^{L_\alpha}   \mathcal{I}[s] \Psi  \frac{\dd{\alpha}}{ B^2 }}{\int_{0}^{L_\alpha} \Psi\frac{\dd{\alpha}}{ B^2 }} .
	\end{dcases}  
\end{align}

Finally, we will study the invertibility of the operator $\Delta_k$ 
\begin{align}
	\Delta_{k} = D_k - U_k \Delta_{k+1}^{-1} L_{k+1} 
	\label{eq:Invertibility_Delta}
\end{align}
assuming that $\Delta_{k+1}$ is an invertible operator from $\mathcal{F}$ to $\mathcal{F}$. For this, first, we note that in the space of functions of interest (smooth periodic functions on the torus), using a Fourier basis $\{e^{\ii (m \theta+ nN_p\zeta)}\}_{m,n\in\mathbb{Z}}$, we can approximate any function $f(\theta,\zeta)=\sum_{m,n\in \mathbb{Z}} \hat{f}_{mn} e^{\ii (m \theta+ nN_p\zeta)} \in \mathcal{F}$ using an approximant $\tilde{f}(\theta,\zeta)$
\begin{align}
	\tilde{f}(\theta,\zeta)=\sum_{- N \le m,n\le N } \hat{f}_{mn} e^{\ii (m \theta+ nN_p\zeta)}
	\label{eq:Fourier_truncated}
\end{align}
truncating the modes with mode number greater than some positive integer $N $ where 
\begin{align}
	\hat{f}_{mn} = \mean*{f,e^{\ii (m \theta+ nN_p\zeta)}}_{\mathcal{F}}  \norm{e^{\ii (m \theta+ nN_p\zeta)}}_{\mathcal{F}}^{-2}
\end{align}
are the Fourier modes of $f$. Thus, we approximate $\mathcal{F}$ using a finite dimensional subspace $\mathcal{F}^{N} \subset \mathcal{F}$ consisting on all the functions of the form given by equation (\ref{eq:Fourier_truncated}).

Hence, we can approximate $D_k$, $U_k$, $\Delta_{k+1}$ and $L_{k+1} $ restricted to $\mathcal{F}^{N}$ (and therefore $\Delta_{k}$) in equation (\ref{eq:Invertibility_Delta}) by operators $D_k^N$, $U_k^N$, $\Delta_{k+1}^N$ and $L_{k+1}^N$ that map any $\tilde{f}\in\mathcal{F}^N$ to the projections of $D_k \tilde{f}$, $U_k \tilde{f}$, $\Delta_{k+1} \tilde{f}$ and $L_{k+1} \tilde{f}$ onto $\mathcal{F}^N$. The operators $D_k^N$, $U_k^N$, $\Delta_{k+1}^N$ and $L_{k+1}^N$ can be exactly represented (in a Fourier basis) by square matrices of size $\dim \mathcal{F}^N$. When the operators are invertible, these matrices are invertible aswell. Doing so, we can interpret the matrix representation of $\Delta_{k}$ as the Schur complement of the matrix
\begin{align}
	M_k^N = 
	\Matrix{cc}
	{ D_k^N & U_k^N \\
		L_{k+1}^N & \Delta_{k+1}^N
	}.
	\label{eq:Invertibility_Delta_Schur}
\end{align}
It is well known from linear algebra that the determinant of $M_k^N$ satisfies
\begin{align}
	\det(M_k^N)
	& =
	\det(\Delta_{k+1}^N)
	\det(\Delta_k^N)
	.
	\label{eq:Schur_complement_determinant}
\end{align}
When both $D_k$ and $\Delta_{k+1}$ are invertible, the matrix $M_k^N$ is invertible. Hence, note from (\ref{eq:Schur_complement_determinant}) that, for $k\ge1$, the matrix $\Delta_{k}^N$ can be inverted for any $N$, and therefore $\Delta_{k}$ (as an operator from $\mathcal{F}$ to $\mathcal{F}$) is invertible.


The case $k=0$ requires special care. In this case $D_0$ is not invertible and the previous argument cannot be applied. In order to make the solution unique, we need to impose an additional constraint to $f^{(0)}$. On ergodic flux-surfaces, condition (\ref{eq:kernel_elimination_condition_Legendre}) is sufficient to fix the value of $f^{(0)}$. However, this is not always the case when $\iota$ is rational. Condition (\ref{eq:kernel_elimination_condition_Legendre}) fixes the value of $f^{(0)}$ solely when the only functions that lie simultaneously at the kernels of $D_0 = -\widehat{E}_\psi \mean*{B^2}^{-1} \vb*{B}\times\nabla\psi\cdot\nabla$ and $L_1=\vb*{b}\cdot\nabla$ are constants (flux-functions). If $\widehat{E}_\psi \ne 0$, this occurs for any $\delta\ne -1/\iota $. However, the case $\delta= -1/\iota$ is unphysical as it would imply $\sqrt{g}=0$. Hence, in practice, when $\widehat{E}_\psi \ne 0$ condition (\ref{eq:kernel_elimination_condition_Legendre}) is sufficient to fix the value of $f^{(0)}$ even if the surface is not ergodic. For rational flux-surfaces and $\widehat{E}_\psi = 0$, condition (\ref{eq:kernel_elimination_condition_Legendre}) is insufficient to fix $f^{(0)}$. In such case, we would need to fix the value of $f^{(0)}$ at a point of each field line as any function $g(\alpha)$ lies in the kernel of $\vb*{b}\cdot\nabla$. In order to clarify this assertion, let's try to obtain $f^{(0)}$ assuming that $f^{(1)}$ is known. Integrating the Legendre mode $k=1$ of equation (\ref{eq:DKE_Forward_elimination}) along a field line gives
\begin{align}
	f^{(0)}(\alpha,l)
	& =
	f^{(0)}_0(\alpha)
	\label{eq:Invertibility_nullspace_condition}
	-
	\int_{0}^{l}
	\left(
	\sigma^{(1)}
	- \Delta_1 f^{(1)} 
	\right)\dd{l'}.
\end{align} 
If $\iota$ is irrational $f_0^{(0)}$ does not depend on $\alpha$. In this case, equation (\ref{eq:Invertibility_nullspace_condition}) and condition (\ref{eq:kernel_elimination_condition_Legendre}) fix $f^{(0)}$ for each $\sigma^{(1)}$, $f^{(1)}$. When $\iota$ is rational we need to distinguish between the case with and without radial electric field. 
\begin{enumerate}
	\item For $E_\psi =0$, the constant $f^{(0)}_0$ is free as no other equation includes $f^{(0)}$. As $f_0^{(0)}$ depends on $\alpha$, condition (\ref{eq:kernel_elimination_condition_Legendre}) does not fix this integration constant. 
	%
	%
	\item For $E_\psi \ne 0$,	
	inserting (\ref{eq:Invertibility_nullspace_condition}) in the Legendre mode $k=0$ of equation (\ref{eq:DKE_Legendre_expansion}) gives
	\begin{align}  
		& -\frac{\widehat{E}_\psi}{\mean*{B^2}}
		B^2 \pdv{f_0^{(0)}}{\alpha} 
		=
		s^{(0)}
		- U_0 f^{(1)}
		\label{eq:Invertibility_Integration_constant_rationals_w_Er}
		\\
		& %
		-
		\frac{\widehat{E}_\psi}{\mean*{B^2}}
		B^2 \pdv{}{\alpha} 
		\int_{0}^{l}
		\left(
		\sigma^{(1)}
		- \Delta_1 f^{(1)} 
		\right)\dd{l'}.\nonumber
	\end{align}
	Integrating $\int_{0}^{L_{\text{c}}} \text{ Eq. (\ref{eq:Invertibility_Integration_constant_rationals_w_Er})}\dd{l}$ gives a differential equation in $\alpha$ from which we can obtain $f_0^{(0)}$ up to a constant. Thus, (\ref{eq:Invertibility_nullspace_condition}), condition (\ref{eq:kernel_elimination_condition_Legendre}) and (\ref{eq:Invertibility_Integration_constant_rationals_w_Er}) fix $f^{(0)}$. 
\end{enumerate} 

Hence, in ergodic flux-surfaces or rational flux-surfaces with finite radial electric field, $M_0^N$ has a one-dimensional kernel. Thus, for $k=0$, it is necessary to substitute one of the rows of $[D_0^N \ \ U_0^N]$ by the condition (\ref{eq:kernel_elimination_condition_Legendre}) so that $M_0^N$ is invertible for any $N$ and as $\Delta_1^N$ can be inverted, also $\Delta_0^N$ constructed in this manner for any $N$, which implies that $\Delta_0$ (as the limit $\lim_{N\rightarrow\infty} \Delta_0^N$) is invertible.
 
 \section{Fourier collocation method} \label{sec:Appendix_Fourier}
 
 In this appendix we describe the Fourier collocation (also called pseudospectral) method for discretizing the angles $\theta$ and $\zeta$. This discretization will be used to obtain the matrices $\vb*{L}_k$, $\vb*{D}_k$ and $\vb*{U}_k$. For convenience, we will use the complex version of the discretization method but for the discretization matrices we will just take their real part as the solutions to (\ref{eq:DKE}) are all real. We search for approximate solutions to equation (\ref{eq:DKE_Legendre_expansion}) of the form
 \begin{align}
 	f^{(k)}(\theta,\zeta) 
 	& = 
 	\sum_{n=-N_{\zeta1}/2}^{N_{\zeta2}/2-1}
 	\sum_{m=-N_{\theta1}/2}^{N_{\theta2}/2-1}
 	\tilde{f}_{mn}^{(k)}
 	e^{\ii(m\theta + nN_{p}\zeta)}
 	\label{eq:Discrete_Fourier_Expansion}
 \end{align}
 where $N_{\theta1} = N_\theta - N_\theta\mod 2 $, $N_{\theta2} = N_\theta + N_\theta\mod 2 $, $N_{\zeta1} = N_\zeta - N_\zeta\mod 2 $, $N_{\zeta2} = N_\zeta + N_\zeta\mod 2 $ for some positive integers $N_\theta$, $N_\zeta$. The complex numbers 
 \begin{align}
 	\tilde{f}_{mn}^{(k)}
 	:=
 	\mean*{f^{(k)}, e^{\ii(m\theta + nN_{p}\zeta)}}_{N_\theta N_\zeta}
 	\norm{ e^{\ii(m\theta + nN_{p}\zeta)}}_{N_\theta N_\zeta}^{-2}
 	\label{eq:Discrete_Fourier_Transform}
 \end{align}
 are the discrete Fourier modes (also called discrete Fourier transform), 
 \begin{align}
 	\mean*{f,g}_{N_\theta N_\zeta}:= 
 	\frac{1}{N_\theta N_\zeta}	
 	\sum_{j'=0}^{N_{\zeta}-1}
 	\sum_{i'=0}^{N_{\theta}-1}
 	f(\theta_{i'},\zeta_{j'})
 	\overline{g(\theta_{i'},\zeta_{j'})}
 	\label{eq:Discrete_Fourier_Inner_product}
 \end{align} 
 is the discrete inner product associated to the equispaced grid points (\ref{eq:Theta_grid}), (\ref{eq:Zeta_grid}), $\norm{ f}_{N_\theta N_\zeta}:=\sqrt{\mean*{f,f}_{N_\theta N_\zeta}}$ its induced norm and $\bar{z}$ denotes the complex conjugate of $z$. We denote by $\mathcal{F}^{N_\theta N_\zeta}$ to the finite dimensional vector space (of dimension $N_\theta N_\zeta$) comprising all the functions that can be written in the form of expansion (\ref{eq:Discrete_Fourier_Expansion}).
 
 The set of functions $\{e^{\ii(m\theta + nN_p\zeta)}\}\subset \mathcal{F}^{N_\theta N_\zeta}$ forms an orthogonal basis for $\mathcal{F}^{N_\theta N_\zeta}$ equipped with the discrete inner product (\ref{eq:Discrete_Fourier_Inner_product}). Namely, 
 \begin{align}
 	\mean*{e^{\ii(m\theta + nN_{p}\zeta)},e^{\ii(m'\theta + n'N_{p}\zeta)}}_{N_\theta N_\zeta} 
 	\propto
 	\delta_{mm'}\delta_{nn'}
 \end{align}
 for $-N_{\theta 1}/2\le m \le N_{\theta 2}/2$ and $-N_{\zeta 1}/2\le n \le N_{\zeta 2}/2$. Thus, for functions lying in $\mathcal{F}^{N_\theta N_\zeta}$, discrete expansions such as (\ref{eq:Discrete_Fourier_Expansion}) coincide with their (finite) Fourier series. The discrete Fourier modes (\ref{eq:Discrete_Fourier_Transform}) are chosen so that the expansion (\ref{eq:Discrete_Fourier_Expansion}) interpolates $f^{(k)}$ at grid points. Hence, there is a vector space isomorphism between the space of discrete Fourier modes and $f^{(k)}$ evaluated at the equispaced grid.

 Combining equations (\ref{eq:Discrete_Fourier_Expansion}), (\ref{eq:Discrete_Fourier_Transform}) and (\ref{eq:Discrete_Fourier_Inner_product}) we can write our Fourier interpolant as
 \begin{align}
 	f^{(k)}(\theta,\zeta) 
 	& = 
 	\vb*{I}(\theta,\zeta) \cdot \vb*{f}^{(k)}
 	\nonumber
 	\\
 	& =
 	\sum_{j'=0}^{N_{\zeta}-1}
 	\sum_{i'=0}^{N_{\theta}-1}
 	I_{i'j'}(\theta,\zeta)
 	f^{(k)}(\theta_{i'},\zeta_{j'})
 	,
 	\label{eq:Fourier_interpolant}
 \end{align}
 where $\vb*{f}^{(k)}\in\mathbb{R}^{N_{\text{fs}}}$ is the state vector containing $f^{(k)}(\theta_{i'},\zeta_{j'})$. The entries of the vector $\vb*{I}(\theta,\zeta)$ are the functions $I_{i'j'}(\theta,\zeta)$ given by, 
 \begin{align}
 	& I_{i'j'}(\theta,\zeta)
 	=
 	I_{i'}^\theta(\theta)
 	I_{j'}^\zeta(\zeta),
 	\\
 	I_{i'}^{\theta}(\theta) &= 
 	\frac{1}{N_\theta}
 	\sum_{m=-N_{\theta1}/2}^{N_{\theta2}/2-1}
 	e^{{ \ii m (\theta-\theta_{i'})} },
 	\\
 	I_{j'}^{\zeta}(\zeta) &= 
 	\frac{1}{N_\zeta}
 	\sum_{n=-N_{\zeta1}/2}^{N_{\zeta2}/2-1}
 	e^{{ N_p\ii n (\zeta-\zeta_{j'})} }
 	.
 \end{align}
 Note that the interpolant is the only function in $\mathcal{F}^{N_\theta N_\zeta}$ which interpolates the data at the grid points, as $I_{i'}^\theta(\theta_i)=\delta_{ii'}$ and $I_{j'}^\zeta(\zeta_j)=\delta_{jj'}$. 
 
 Of course, our approximation (\ref{eq:Fourier_interpolant}) cannot (in general) be a solution to (\ref{eq:DKE_Legendre_expansion}) at all points $(\theta,\zeta)\in[0,2\pi)\times[0,2\pi/N_p)$. Instead, we will force that the interpolant (\ref{eq:Fourier_interpolant}) solves equation (\ref{eq:DKE_Legendre_expansion}) exactly at the equispaced grid points. Thanks to the vector space isomorphism (\ref{eq:Discrete_Fourier_Transform}) between $\vb*{f}^{(k)}$ and the discrete modes $\tilde{f}_{mn}^{(k)}$ this is equivalent to matching the discrete Fourier modes of the left and right-hand-sides of equation (\ref{eq:DKE_Legendre_expansion}).
 
 Inserting the interpolant (\ref{eq:Fourier_interpolant}) in the left-hand side of equation (\ref{eq:DKE_Legendre_expansion}) and evaluating the result at grid points gives
 \begin{align}
 	& 
 	\eval{\left(
 		L_k f^{(k-1)} 
 		+
 		D_k f^{(k)}
 		+
 		U_k f^{(k+1)}
 		\right)}_{(\theta_i,\zeta_j)}
 	=
 	\nonumber
 	\\
 	& 
 	\eval{\left(
 		L_k \vb*{I} \cdot \vb*{f}^{(k-1)} 
 		+
 		D_k \vb*{I} \cdot \vb*{f}^{(k)}
 		+
 		U_k \vb*{I} \cdot \vb*{f}^{(k+1)}
 		\right)}_{(\theta_i,\zeta_j)}.
 \end{align}
 Here, $L_k \vb*{I}(\theta_i,\zeta_j)$, $D_k \vb*{I}(\theta_i,\zeta_j)$ and $U_k \vb*{I}(\theta_i,\zeta_j)$ are respectively the rows of $\vb*{L}_k$, $\vb*{D}_k$ and $\vb*{U}_k$ associated to the grid point $(\theta_i,\zeta_j)$. We can relate them to the actual positions they will occupy in the matrices choosing an ordenation of rows and columns. We use the ordenation that relates respectively the row $i_{\text{r}}$ and column $i_{\text{c}}$ to the grid points $(\theta_i,\zeta_j)$ and $(\theta_{i'},\zeta_{j'})$ as
 \begin{align}
 	i_{\text{r}} & = 1 + i + j N_\theta,  \label{eq:Row_ordenation}\\ 
 	i_{\text{c}} & = 1 + i' + j' N_\theta, \label{eq:Column_ordenation}
 \end{align}
 for $i,i'=0,1,\ldots,N_\theta-1$ and  $j,j'=0,1,\ldots,N_\zeta-1$. With this ordenation, we define the elements of the row $i_{\text{r}}$ and column $i_{\text{c}}$ given by (\ref{eq:Row_ordenation}) and (\ref{eq:Column_ordenation}) of the matrices $\vb*{L}_k$, $\vb*{D}_k$ and $\vb*{U}_k$ to be 
 \begin{align}
 	\left(\vb*{L}_k\right)_{i_{\text{r}} i_{\text{c}}}
 	& =
 	{L_k I_{i'j'}}{(\theta_i,\zeta_j)},
 	\\
 	\left(\vb*{D}_k\right)_{i_{\text{r}} i_{\text{c}}}
 	& =
 	{D_k I_{i'j'}}{(\theta_i,\zeta_j)},
 	\\
 	\left(\vb*{U}_k\right)_{i_{\text{r}} i_{\text{c}}}
 	& =
 	{U_k I_{i'j'}}{(\theta_i,\zeta_j)}.
 \end{align}
 Explicitly,
 \begin{align}
 	\eval{L_k I_{i'j'}}_{(\theta_i,\zeta_j)}
 	& =
 	\frac{k}{2k-1} 
 	\left(
 	\eval{\vb*{b} \cdot \nabla I_{i'j'}}_{(\theta_i,\zeta_j)}
 	\nonumber
 	\right.
 	\\
 	&
 	+
 	\frac{k-1}{2}
 	\left.
 	\eval{\vb*{b}\cdot\nabla \ln B}_{(\theta_i,\zeta_j)}	
 	\delta_{ii'}\delta_{jj'}
 	\right)
 	,
 	\\
 	\eval{D_k I_{i'j'}}_{(\theta_i,\zeta_j)}
 	& =
 	-\frac{\widehat{E}_\psi}{\mean*{B^2}}
 	\eval{\vb*{B}\times \nabla\psi  \cdot \nabla 
 		I_{i'j'}}_{(\theta_i,\zeta_j)}
 	\nonumber \\
 	& +  
 	\frac{k(k+1)}{2}
 	\hat{\nu}\delta_{ii'}\delta_{jj'}
 	,
 	\\
 	\eval{U_k I_{i'j'}}_{(\theta_i,\zeta_j)}
 	& = 
 	\frac{k+1}{2k+3} 
 	\left(
 	\eval{\vb*{b} \cdot \nabla  I_{i'j'}}_{(\theta_i,\zeta_j) } 
 	\right. \nonumber
 	\\
 	& +
 	\left.
 	\frac{k+2}{2}
 	\eval{\vb*{b}\cdot\nabla \ln B}_{(\theta_i,\zeta_j)}	
 	\delta_{ii'}\delta_{jj'}
 	\right)
 	,
 \end{align}
 where we have used expressions (\ref{eq:Parallel_streaming_spatial_operator}) and (\ref{eq:ExB_spatial_operator}) to write
 \begin{align}
 	& \eval{\vb*{b} \cdot \nabla  I_{i'j'}}_{(\theta_i,\zeta_j) }
 	=
 	\eval{\frac{B}{B_\zeta + \iota B_\theta}}_{(\theta_i,\zeta_j)}
 	\nonumber\\
 	& \qquad \times
 	\left(
 	\iota 
 	\delta_{jj'}
 	\eval{\dv{I_{i'}^{\theta}}{\theta}}_{\theta_i}
 	\right.
 	-
 	\left.
 	\delta_{ii'}
 	\eval{\dv{I_{j'}^{\zeta}}{\zeta}}_{\zeta_j}
 	\right),
 	\\
 	& \eval{\vb*{B}\times \nabla\psi  \cdot \nabla 
 		I_{i'j'}}_{(\theta_i,\zeta_j)}
 	=
 	\eval{\frac{B^2}{B_\zeta + \iota B_\theta}}_{(\theta_i,\zeta_j)}
 	\nonumber \\ 
 	&
 	\qquad
 	\times\left(
 	B_\zeta 
 	\delta_{jj'}
 	\eval{\dv{I_{i'}^{\theta}}{\theta}}_{\theta_i}
 	\right. 
 	-
 	\left.
 	B_\theta 
 	\delta_{ii'}
 	\eval{\dv{I_{j'}^{\zeta}}{\zeta}}_{\zeta_j}
 	\right).
 \end{align}
 We remark that, for $k=0$, the rows of $\vb*{D}_0$ and $\vb*{U}_0$ associated to the grid point $(\theta_0,\zeta_0)=(0,0)$, are replaced by equation (\ref{eq:kernel_elimination_condition_Legendre}). Finally, each state vector $\vb*{f}^{(k)}$ for the Fourier interpolants contains the images $f^{(k)}(\theta_{i'},\zeta_{j'})$ at the grid points, ordered according to (\ref{eq:Column_ordenation}).

\section{Convergence of monoenergetic coefficients calculated by {\DKES}}\label{sec:Appendix_DKES_Bounds}

\FloatBarrier
The code {\DKES} gives an approximation to the monoenergetic geometric coefficients as a semisum of two quantities $\widehat{D}_{ij}^- $ and $\widehat{D}_{ij}^+$ by solving a variational principle \cite{VanRij_1989}. For each coefficient, the output of {\DKES} consists on two quantities $\widehat{D}_{ij}^\mp K_{ij}$, where $K_{ij}$ are the normalization factors
\begin{align}
	K_{ij} & :=\left(\dv{\psi}{r}\right)^{-2}, &\quad i,j \in\{1,2\},
	\\
	K_{i3} & :=  \left(\dv{\psi}{r} \right)^{-1}, &\quad i \in\{1,2\},
	\\
	K_{3j} & := \left(\dv{\psi}{r} \right)^{-1}, &\quad j \in\{1,2\},
	\\
	K_{33} & := 1, &
\end{align} 
to change from the radial coordinate $\psi$ to $r$. In table \ref{tab:DKES_normalization_factors}, the normalization factors for the configurations considered are listed. 
\begin{table}[h]
	\centering
	\begin{tabular}{@{}lccc@{}}
		\toprule
		Configuration & $\dv*{\psi}{r}$ & $K_{11}$    & $K_{31}$    
		\\ \midrule
		W7X-EIM       & 0.5237  & 3.6462 & 1.9095 \\
		W7X-KJM       & 0.5132  & 3.7969 & 1.9486 \\
		CIEMAT-QI     & 0.4674  & 4.5774 & 2.1395 \\ \bottomrule
	\end{tabular}
	\caption{Normalization factors for {\DKES} results. $\dv*{\psi}{r}$ in $\text{T}\cdot\text{m}$, $K_{11}$ in $\text{T}^{-2}\cdot\text{m}^{-2}$ and $K_{31}$ in $\text{T}^{-1}\cdot\text{m}^{-1}$.}
	\label{tab:DKES_normalization_factors}
\end{table}

Apart from the normalization factors, there is still a nuance left for the parallel conductivity coefficient: the code {\DKES} computes this coefficient measured with respect to the one obtained by solving the Spitzer problem
\begin{align}
	-\hat{\nu} \Lorentz f_{\text{Sp}} = s_3.
\end{align}
Using (\ref{eq:Legendre_eigenvalues}) is immediate to obtain the $1-$th Legendre mode of $f_{\text{Sp}}$
\begin{align}
	f_{\text{Sp}}^{(1)} =  \frac{1}{\hat{\nu}} \frac{B}{B_0}
\end{align}
and using (\ref{eq:Gamma_33_Legendre}) we obtain its associated $\widehat{D}_{33}$ coefficient
\begin{align}
	\widehat{D}_{33,\text{Sp}} & = \frac{2}{3\hat{\nu}} \mean*{\frac{B^2}{B_0^2}}.
\end{align}
Thus, the output of {\DKES} for the parallel conductivity coefficient has to be compared against the deviation $(\widehat{D}_{33} - \widehat{D}_{33,\text{Sp}})$. 

From the output of {\DKES}, the diagonal elements $\widehat{D}_{ii}^{\pm}$ satisfy $\widehat{D}_{ii}^{-} \ge \widehat{D}_{ii} \ge\widehat{D}_{ii}^{+}$ and allow to compute bounds for $\widehat{D}_{ij}$
\begin{align}
	\frac{\widehat{D}_{ij}^{-} + \widehat{D}_{ij}^{+}}{2}
	-
	\Delta_{ij}
	\le
	\widehat{D}_{ij}
	\le
	\frac{\widehat{D}_{ij}^{-} + \widehat{D}_{ij}^{+}}{2}
	+
	\Delta_{ij}
\end{align}
and $\Delta_{ij} = \sqrt{(\widehat{D}_{ii}^{-} - \widehat{D}_{ii}^{+})(\widehat{D}_{jj}^{-} - \widehat{D}_{jj}^{+})} /2 $.  

\begin{figure}[t]
	\centering	
	\begin{subfigure}[t]{0.32\textwidth}
		\tikzsetnextfilename{DKES-Convergence-Legendre-W7X-EIM-s0200-Er-0-D31-Detail}
		\includegraphics{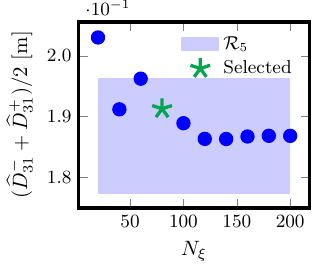}
		\caption{}
		\label{subfig:DKES_D31_convergence_Legendre_W7X_EIM_0200_Erho_0_Detail}
	\end{subfigure}
	\begin{subfigure}[t]{0.32\textwidth}
		\tikzsetnextfilename{DKES-Convergence-theta-zeta-W7X-EIM-s0200-Er-0-D31}
		\includegraphics{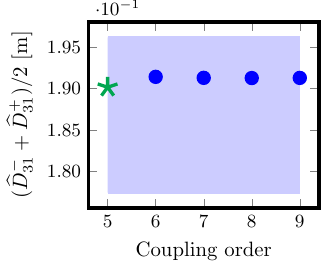}
		\caption{}
		\label{subfig:DKES_D31_convergence_Coupling_parameter_W7X_EIM_0200_Erho_0}
	\end{subfigure}

	\caption{Convergence of $(\widehat{D}_{31}^- + \widehat{D}_{31}^+) /2$ computed with {\DKES} for W7X-EIM at the surface labelled by $\psi/\psi_{\text{lcfs}}=0.200$, for $\hat{\nu}(v)=10^{-5}$ $\text{m}^{-1}$ and $\widehat{E}_r(v)=0$ $\text{kV}\cdot\text{s}/\text{m}^2$. (a) Convergence with $N_\xi$ for coupling order = 9. (b) Convergence with the coupling order for $N_\xi= 80$.}
	\label{fig:DKES_Convergence_W7X_EIM_Er_0}
\end{figure}

\begin{figure}[t]
	\centering
	\begin{subfigure}[t]{0.32\textwidth}
		\tikzsetnextfilename{DKES-Convergence-Legendre-W7X-EIM-s0200-Er-3e-4-D31-Detail}
		\includegraphics{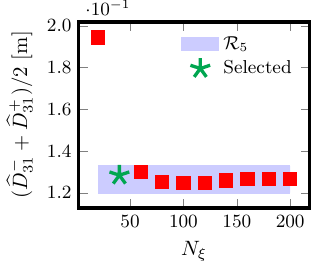}
		\caption{}
		\label{subfig:DKES_D31_convergence_Legendre_W7X_EIM_0200_Erho_3e-4_Detail}
	\end{subfigure}
	\begin{subfigure}[t]{0.32\textwidth}
		\tikzsetnextfilename{DKES-Convergence-theta-zeta-W7X-EIM-s0200-Er-3e-4-D31}
		\includegraphics{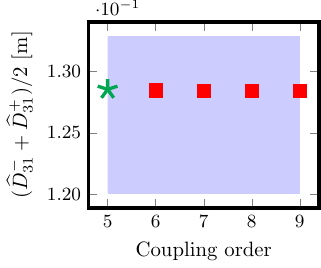}
		\caption{}
		\label{subfig:DKES_D31_convergence_Coupling_parameter_W7X_EIM_0200_Erho_3e-4}
	\end{subfigure}
	
	\caption{Convergence of $(\widehat{D}_{31}^- + \widehat{D}_{31}^+) /2$ computed with {\DKES} for W7X-EIM at the surface labelled by $\psi/\psi_{\text{lcfs}}=0.200$, for $\hat{\nu}(v)=10^{-5}$ $\text{m}^{-1}$ and $\widehat{E}_r(v)=3\cdot 10^{-4}$ $\text{kV}\cdot\text{s}/\text{m}^2$. (a) Convergence with $N_\xi$ for coupling order = 9. (b) Convergence with the coupling order for $N_\xi= 40$.}
	\label{fig:DKES_Convergence_W7X_EIM_Er_3e-4}
\end{figure}

\begin{figure}[t]
	\centering	
	\begin{subfigure}[t]{0.32\textwidth}
		\tikzsetnextfilename{DKES-Convergence-Legendre-W7X-KJM-s0204-Er-0-D31-Detail}
		\includegraphics{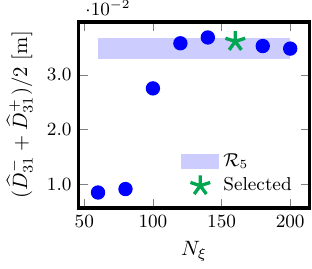}
		\caption{}
		\label{subfig:DKES_D31_convergence_Legendre_W7X_KJM_0204_Erho_0_Detail}
	\end{subfigure}
	\begin{subfigure}[t]{0.32\textwidth}
		\tikzsetnextfilename{DKES-Convergence-theta-zeta-W7X-KJM-s0204-Er-0-D31}
		\includegraphics{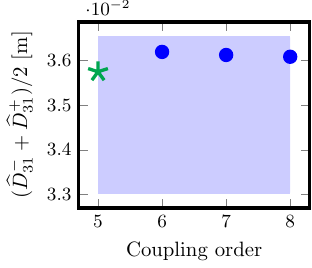}
		\caption{}
		\label{subfig:DKES_D31_convergence_Coupling_parameter_W7X_KJM_0204_Erho_0}
	\end{subfigure}

	\caption{Convergence of $(\widehat{D}_{31}^- + \widehat{D}_{31}^+) /2$ computed with {\DKES} for W7X-KJM at the surface labelled by $\psi/\psi_{\text{lcfs}}=0.204$, for $\hat{\nu}(v)=10^{-5}$ $\text{m}^{-1}$ and $\widehat{E}_r(v)=0$ $\text{kV}\cdot\text{s}/\text{m}^2$. (a) Convergence with $N_\xi$ for coupling order = 8. (b) Convergence with the coupling order for $N_\xi=160 $.}
	\label{fig:DKES_Convergence_W7X_KJM_Er_0}
\end{figure}

\begin{figure}[t]
	\centering
	\begin{subfigure}[t]{0.32\textwidth}
		\tikzsetnextfilename{DKES-Convergence-Legendre-W7X-KJM-s0204-Er-3e-4-D31-Detail}
		\includegraphics{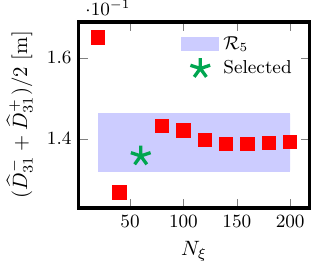}
		\caption{}
		\label{subfig:DKES_D31_convergence_Legendre_W7X_KJM_0204_Erho_3e-4_Detail}
	\end{subfigure}
	\begin{subfigure}[t]{0.32\textwidth}
		\tikzsetnextfilename{DKES-Convergence-theta-zeta-W7X-KJM-s0204-Er-3e-4-D31}
		\includegraphics{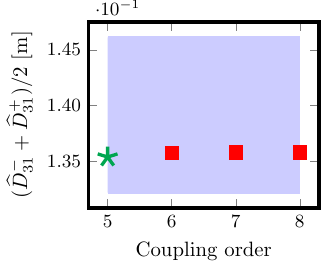}
		\caption{}
		\label{subfig:DKES_D31_convergence_Coupling_parameter_W7X_KJM_0204_Erho_3e-4}
	\end{subfigure}
	
	\caption{Convergence of $(\widehat{D}_{31}^- + \widehat{D}_{31}^+) /2$ computed with {\DKES} for W7X-KJM at the surface labelled by $\psi/\psi_{\text{lcfs}}=0.204$, for $\hat{\nu}(v)=10^{-5}$ $\text{m}^{-1}$ and $\widehat{E}_r(v)=3\cdot 10^{-4}$ $\text{kV}\cdot\text{s}/\text{m}^2$. (a) Convergence with $N_\xi$ for coupling order = 7. (b) Convergence with the coupling order for $N_\xi= 60$.}
	\label{fig:DKES_Convergence_W7X_KJM_Er_3e-4}
\end{figure}

\begin{figure}[t]
	\centering	
	\begin{subfigure}[t]{0.32\textwidth}
		\tikzsetnextfilename{DKES-Convergence-Legendre-CIEMAT-QI-s0250-Er-0-D31-Detail}
		\includegraphics{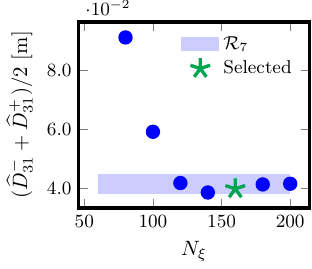}
		\caption{}
		\label{subfig:DKES_D31_convergence_Legendre_CIEMAT_QI_0250_Erho_0_Detail}
	\end{subfigure}
	\begin{subfigure}[t]{0.32\textwidth}
		\tikzsetnextfilename{DKES-Convergence-theta-zeta-CIEMAT-QI-s0250-Er-0-D31}
		\includegraphics{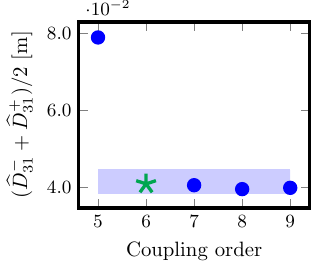}
		\caption{}
		\label{subfig:DKES_D31_convergence_Coupling_parameter_CIEMAT_QI_0250_Erho_0}
	\end{subfigure}

	\caption{Convergence of $(\widehat{D}_{31}^- + \widehat{D}_{31}^+) /2$ computed with {\DKES} for CIEMAT-QI at the surface labelled by $\psi/\psi_{\text{lcfs}}=0.250$, for $\hat{\nu}(v)=10^{-5}$ $\text{m}^{-1}$ and $\widehat{E}_r(v)=0$ $\text{kV}\cdot\text{s}/\text{m}^2$. (a) Convergence with $N_\xi$ for coupling order = 9. (b) Convergence with the coupling order for $N_\xi= 160$.}
	\label{fig:DKES_Convergence_CIEMAT_QI_Er_0}
\end{figure}

\begin{figure}[t]
	\centering	
	\begin{subfigure}[t]{0.32\textwidth}
		\includegraphics{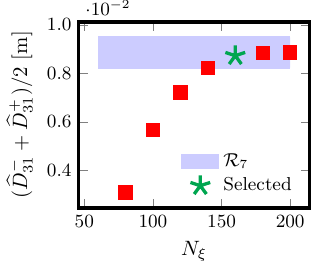}
		\caption{}
		\label{subfig:DKES_D31_convergence_Legendre_CIEMAT_QI_0250_Erho_1e-3_Detail}
	\end{subfigure}
	\begin{subfigure}[t]{0.32\textwidth}
		\tikzsetnextfilename{DKES-Convergence-theta-zeta-CIEMAT-QI-s0250-Er-1e-3-D31}
		\includegraphics{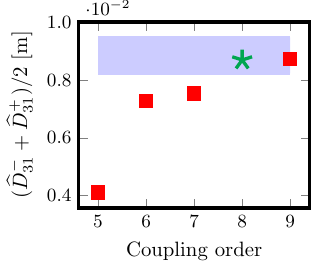}
		\caption{}
		\label{subfig:DKES_D31_convergence_Coupling_parameter_CIEMAT_QI_0250_Erho_1e-3}
	\end{subfigure}

	\caption{Convergence of $(\widehat{D}_{31}^- + \widehat{D}_{31}^+) /2$ computed with {\DKES} for CIEMAT-QI at the surface labelled by $\psi/\psi_{\text{lcfs}}=0.250$, for $\hat{\nu}(v)=10^{-5}$ $\text{m}^{-1}$ and $\widehat{E}_r(v)=10^{-3}$ $\text{kV}\cdot\text{s}/\text{m}^2$. (a) Convergence with $N_\xi$ for coupling order = 9. (b) Convergence with the coupling order for $N_\xi= 160$.}
	\label{fig:DKES_Convergence_CIEMAT_QI_Er_1e-3}
\end{figure}

In figures \ref{fig:DKES_Convergence_W7X_EIM_Er_0}, \ref{fig:DKES_Convergence_W7X_EIM_Er_3e-4}, \ref{fig:DKES_Convergence_W7X_KJM_Er_0}, \ref{fig:DKES_Convergence_W7X_KJM_Er_3e-4}, \ref{fig:DKES_Convergence_CIEMAT_QI_Er_0} and \ref{fig:DKES_Convergence_CIEMAT_QI_Er_1e-3} the convergence study for selecting {\DKES} resolutions is shown. In the code {\DKES} the number of Legendre modes used are specified by $N_\xi$. In order to select the number of Fourier modes in the Boozer angles $(\theta,\zeta)$ that {\DKES} uses, an integer called ``coupling order'' must be specified. Using figures \ref{subfig:DKES_D31_convergence_Legendre_W7X_EIM_0200_Erho_0_Detail}, \ref{subfig:DKES_D31_convergence_Legendre_W7X_EIM_0200_Erho_3e-4_Detail}, \ref{subfig:DKES_D31_convergence_Legendre_W7X_KJM_0204_Erho_0_Detail}, \ref{subfig:DKES_D31_convergence_Legendre_W7X_KJM_0204_Erho_3e-4_Detail}, \ref{subfig:DKES_D31_convergence_Legendre_CIEMAT_QI_0250_Erho_0_Detail} and \ref{subfig:DKES_D31_convergence_Legendre_CIEMAT_QI_0250_Erho_1e-3_Detail}, the number of Legendre modes $N_\xi$ is selected so that it satisfies convergence condition (i) using the region $\mathcal{R}_\epsilon$ for each case. After that, using \ref{subfig:DKES_D31_convergence_Coupling_parameter_W7X_EIM_0200_Erho_0}, \ref{subfig:DKES_D31_convergence_Coupling_parameter_W7X_EIM_0200_Erho_3e-4}, \ref{subfig:DKES_D31_convergence_Coupling_parameter_W7X_KJM_0204_Erho_0}, \ref{subfig:DKES_D31_convergence_Coupling_parameter_W7X_KJM_0204_Erho_3e-4}, \ref{subfig:DKES_D31_convergence_Coupling_parameter_CIEMAT_QI_0250_Erho_0} and \ref{subfig:DKES_D31_convergence_Coupling_parameter_CIEMAT_QI_0250_Erho_1e-3}, we select the minimum value of the coupling order for which the calculation with the selected value of $N_\xi$ satisfies convergence condition (ii).

\FloatBarrier
\section*{References}
\bibliographystyle{unsrt}
\bibliography{refs}{}

\end{document}